# Nonlinear parametric excitation effect induces stability transitions in swimming direction of flexible superparamagnetic microswimmers

Yuval Harduf[1†], Dongdong Jin[2†], Yizhar Or[1*] and Li Zhang[2,3*]

**Abstract:**

Microscopic artificial swimmers have recently become highly attractive due to their promising potential for biomedical applications. The pioneering work of Dreyfus *et al* (2005) have demonstrated the motion of a microswimmer with an undulating chain of superparamagnetic beads, which is actuated by applying an oscillating external magnetic field. Interestingly, they have shown that the swimmer's orientation undergoes a 90°-transition when the magnetic field's oscillations amplitude is increased above a critical value. In this work, we further investigate this transitions both theoretically and experimentally by using numerical simulations and presenting a novel flexible microswimmer with a superparamagnetic head. We prove that this effect depends on both frequency and amplitude of the oscillating magnetic field, and demonstrate existence of an optimal amplitude achieving maximal swimming speed. Asymptotic analysis of a minimal two-link model reveals that the changes in the swimmer's direction represent stability transitions which are induced by a nonlinear parametric excitation.

[1]Faculty of Mechanical Engineering, Technion – Israel Institute of Technology, Haifa 3200003, Israel.

[2]Department of Mechanical and Automation Engineering, The Chinese University of Hong Kong, Shatin NT, Hong Kong SAR, China.

[3]Chow Yuk Ho Technology Center for Innovative Medicine, The Chinese University of Hong Kong, Shatin NT, Hong Kong SAR, China.

[†]These authors contributed equally to this work.

[*]Corresponding author, email: izi@me.technion.ac.il and lizhang@mae.cuhk.edu.hk

## Introduction

Remotely controlled micro-robotic swimmers have recently become highly attractive in the engineering science community, due to their promising potential for biomedicine[1,2]. Drawing biological inspiration from swimming microorganisms[3,4], micro-robotic swimmers are planned to be utilized for in-body biomedical tasks such as targeted drug delivery[5,6], therapeutic diagnosis, and even assisting sperm's motility[7]. The motion of microswimmers is dominated by viscous drag forces while inertial effects are negligible[8,9]. A common actuation method of microswimmers is applying time-varying external magnetic fields[10,11]. Several works presented rigid helical microswimmers[12-15] actuated by a rotating magnetic field which induces corkscrew-like swimming. Dreyfus *et al*[10] presented a microswimmer composed of a flexible chain of superparamagnetic beads connected to a red blood cell, where a spatially uniform magnetic field **B** varies in time as

$$\mathbf{B}(t) = B_x \hat{\mathbf{x}} + B_y \sin(2\pi f t)\hat{\mathbf{y}} = B_x(\hat{\mathbf{x}} + \beta \sin(2\omega t)\hat{\mathbf{y}}). \tag{1}$$

In a later work[16], it has been shown that when the amplitude ratio $\beta = B_y/B_x$ is increased beyond a critical value of $\sqrt{2}$, microswimmers composed of a chain of beads reached a sharp 90°-transition in their mean orientation. The "ineffective" swimmers studied in[16] did not display any net swimming, since they were not attached to a load that is needed for breaking the swimmer's front-back symmetry. The 90°-transition has also been confirmed by numerical simulations[17] for a microswimmer with a load, which also displayed a change in the swimming direction from *x* to *y*. Nevertheless, no physical explanation has been provided to this effect.

In this work, we further study this transition both theoretically and experimentally. Unlike previous works[16,17], we find that the conditions for transition can depend on both the ratio β and the frequency *f*. For a fixed frequency, we also find an optimal value of β which maximizes the swimming speed in *y* direction. These results are confirmed by using numerical simulations of a multi-link superparamagnetic microswimmer, and by presenting experimental results of a novel elastic microswimmer with a superparamagnetic head that breaks front-back symmetry. Next, we provide an analytic explanation to this transition by presenting a minimal theoretical model of a two-link swimmer with an elastic joint. We explicitly formulate the swimmer's dynamics as a low-dimensional nonlinear system with parametric excitation, which induces transitions in the stability of different periodic solutions, corresponding to oscillations about and swimming along *x* or *y* directions. In the limit of fast oscillations and low stiffness, asymptotic analysis using the method of multiple scales reveals the previously observed transition at β=√2. Approximate expressions for the mean swimming speed are obtained, which clearly indicate existence of an optimal value of β that attains maximal speed. In the limit of fast oscillations and high stiffness, asymptotic approximation of the system yields a scalar second-order differential equation with parametric excitation[18], which resembles the well-known Kapitza pendulum[19]. Analysis of stability transitions in this system gives conditions that depend on both the ratio β and the frequency *f*, in agreement with our experiments and simulation results.

## Numerical simulations

We now present our numerical simulations of a multi-link microswimmer. A schematic description of the swimmer appears in Fig. 1a. It consists of a chain of $N=10$ slender rigid links connected by rotary joints, where each link carries a superparamagnetic bead. The chain, whose total length is $L$, and is attached to a spherical head which represents the load. The external magnetic field undergoes planar oscillations as described in (1). A detailed derivation of the microswimmer's equations of motion appears in the Supplementary Information (SI), and key points are briefly summarized here. Elasticity of the chain is represented by torsion springs, such that the elastic bending torque acting on each joint is given by $\tau_i = -k\phi_i$, where $\phi_i$ is the relative angle at joint $i$. The magnetic field induces an external torque $\mathbf{L}_i$ on each link, which is dictated by the interaction between dipoles generated by the superparamagnetic beads. For calculating viscous drag forces acting on the swimmer, we use Stokes drag for a sphere[7] and *resistive force theory*[20,21] for the slender links, while hydrodynamic interaction is neglected. All this gives a nonlinear system of first-order differential equations which are integrated numerically (see SI). We calibrate our system parameters using data from the experiments of Dreyfus *et al*[10], and then simulate the microswimmer's motion under a moderate amplitude ratio of $\beta \approx 1$ for different values of actuation frequency $f$ of the magnetic field as given in (1). We find that the swimmer's net motion is along $x$ direction, and that a maximal speed is obtained at an optimal value of the frequency. The results of these simulations show good agreement with those reported by Dreyfus *et al*[10], as shown in Fig. 1b.

Next, we verify the results given by Roper *et al*[16], by considering an elastic chain without a load, composing an "ineffective swimmer". For a fixed actuation frequency $f$ of the magnetic field in (1) and a fixed magnitude of the constant component of magnetic field $B_x$, we gradually increase the amplitude of the oscillating component $B_y$, until reaching a critical value where the swimmer's orientation converges in steady state to oscillations about $y$ rather than $x$ direction. For a frequency of $f=50Hz$, the critical value of $B_y$ as a function of $B_x$ is shown in Fig. 1c, showing a nearly constant critical ratio of $\beta_{cr}=\sqrt{2}$, in agreement with Roper *et al*[16]. However, when the actuation frequency is decreased to $f=5Hz$, a noticeable change occurs. As shown in Fig. 1c, the value of $\beta_{cr}$ now increases significantly above $\sqrt{2}$ for large values of $B_x$. This proves that the conditions for transition in the swimmer's direction depend on frequency $f$ and the constant magnetic field $B_x$, in addition to the amplitude ratio $\beta$.

In order to show a change in the swimming direction and not just mean orientation of the swimmer's oscillations, we conducted additional simulations for Dreyfus' microswimmer with a spherical load. Fig. 1d shows the mean swimming speed $V$ as a function of the actuation frequency $f$ for a constant amplitude ratio of $\beta=1.6$, clearly showing a transition from swimming in $x$ direction to $y$ direction at $f \approx 1.8Hz$. This proves again that this transition strongly depends on the actuation frequency $f$, in addition to $\beta$. Finally, we simulate the microswimmer's motion under a constant frequency $f=20Hz$, and plot the speed $V$ as a function of $\beta$, in Fig. 1e. The results show that the speed depends non-monotonically on $\beta$, and that an optimal value of $\beta$ exists, which maximizes the swimming speed. Remarkably, this maximal speed is obtained beyond the transition to swimming in $y$ direction.

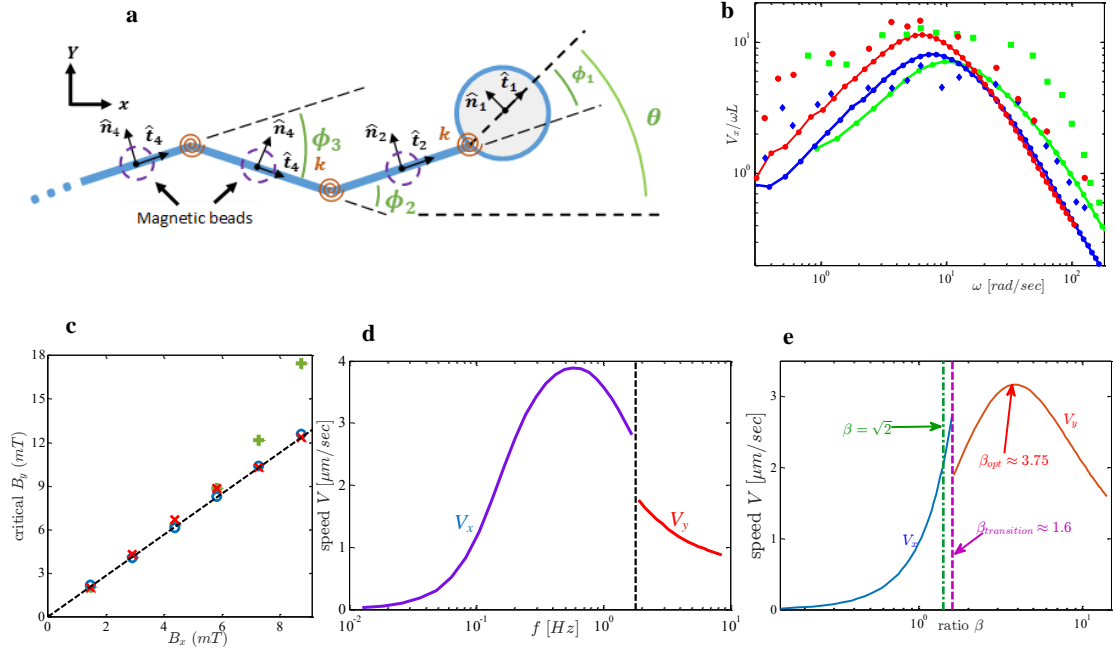

**Figure 1:** (a) The planar multi-link microswimmer model with a non-magnetic spherical load. The rigid slender links carry small superparamagnetic beads (dashed circles), and the connecting joints contain torsion springs. (b) Normalized swimming speed $V_x/\omega L$ as a function of actuation frequency $\omega$ (logarithmic scale). Comparison of our numerical simulations (solid lines) with Dreyfus' experimental data[10] (discrete markers). Parameter values are given in the SI. (c) Critical value of the magnetic field's oscillating component $B_y=\beta B_x$ for transition from oscillations about $x$ direction to $y$ direction, as a function of the constant component $B_x$. The 'x' markers are from the experiment of Roper *et al*[16] under frequency $f$=50Hz. The 'o' markers are our simulations of a multi-link swimmer without a spherical load for $f$=50Hz. The '+' markers are our simulations of a multi-link swimmer without a spherical load for $f$=5Hz. The dashed line is of slope $\sqrt{2}$. It can be seen that the critical value of $\beta=B_y/B_x$ depends also on $f$ and $B_x$, and differs from $\sqrt{2}$. (d) Numerical simulations of the mean speed $V$ of Dreyfus' microswimmer with a spherical load as a function of actuation frequency $f$ (logarithmic scale) for $B_y=1.6B_x=13.92mT$. The dashed vertical line denotes critical frequency where the steady-state solution changes from swimming in $x$ direction to $y$ direction. The labels $V_x$ and $V_y$ denote swimming speed in $x$ and $y$ directions, respectively. (e) Numerical simulations of the mean speed $V$ of Dreyfus' microswimmer with a spherical load as a function of amplitude ratio $\beta$ (logarithmic scale) for $B_x=8.7mT$ and $f=10Hz$. The dashed vertical line denotes critical value of $\beta\approx1.6$ where the steady-state solution changes from swimming in $x$ direction to $y$ direction. It is noticeably larger than $\beta = \sqrt{2}$ (dash-dotted line). The labels $V_x$ and $V_y$ denote swimming speed in $x$ and $y$ directions, respectively. It can be seen that there exist an optimal value of $\beta$ that maximizes swimming speed $V_y$, which is greater than $V_x$.

**Experiment**

We now present experimental results with a newly fabricated microswimmer in order to verify our findings on the transition in the swimming direction. Our experimental microswimmer prototype consists of two parts: a superparamagnetic head actuated by an external oscillating magnetic field, and a flexible main body which enables breaking the time-reversibility of motion during body undulations. The main body of the microswimmer is made of polypyrrole (Ppy), a kind of conductive and elastic polymer. The superparamagnetic head was made by embedding $Fe_3O_4$ nanoparticles ($Fe_3O_4$ NPs) at the end of the Ppy filament. The fabrication process, including preparation of superparamagnetic $Fe_3O_4$ NPs, electrophoretic codeposition of $Fe_3O_4$ NPs and Ppy followed by electrochemical deposition of Ppy in an anodic aluminum oxide (AAO) template, is shown in Fig. 2 (Detailed information regarding all the experimental procedures are available in the SI). It is worth mentioning that the average diameter and zeta potential of $Fe_3O_4$ NPs prepared by hydrothermal method in electrophoresis bath was approximately 80 nm and -50 mV, respectively (shown in Fig. S1(b) and (d)), demonstrating that they could be assembled into AAO template with 200 nm-diameter pores by electrophoretic deposition. Typical SEM images (see Fig. S1(e) and (f) in the SI) depict the morphology of as-prepared microswimmers, and it is found that the typical length and diameter of the microswimmers are 25-30 μm and approximately 200 nm, respectively.

The microswimmer is actuated by a planar oscillating magnetic field generated using a 3-axis Helmholtz electromagnetic coil setup. The generated planar oscillating field **B**(*t*) was a spatially uniform time-varying external magnetic field, having a constant component of $B_x$=5mT in *x* direction and a sinusoidally oscillating component of $b\beta\sin(2\pi ft)$ in *y* direction, as described in Eq. (1).

Under small-amplitude oscillation of the magnetic field where the ratio *β* is as low as 0.5, the swimming behavior of the microswimmer is very close to previously reported flexible magnetic microswimmers[10,11]. For better understanding, Fig. 2(a) shows a sequence of pictures illustrating the dynamic motion of microswimmer during a 20-second duration under an oscillating magnetic field with the input parameters of $\beta$=0.5 and $f$=10 Hz. As expected, the superparamagnetic head of the microswimmer continuously tends to align with the direction of the magnetic field, resulting in undulations of the Ppy flexible main body. As a consequence, the microswimmer oscillates about and swims along the mean direction of the magnetic field, $x$ axis, with a translational speed of $V_x$=0.69 μm/s. Repeating this process for varied frequencies $f$, our experimental results indicate a resonance-like dependence of swimming velocity on $f$ as shown in Fig. 2(b). That is, the maximal speed $V_x$ is attained at an optimal input frequency. This resonance-like effect is in agreement with our numerical simulations and the results of Dreyfus *et al*[10]. Next, when the amplitude ratio $\beta$ is increased to 1, the dynamic motion and average speed of the microswimmer as a function of frequency are shown in Fig. 2(c) and (d).

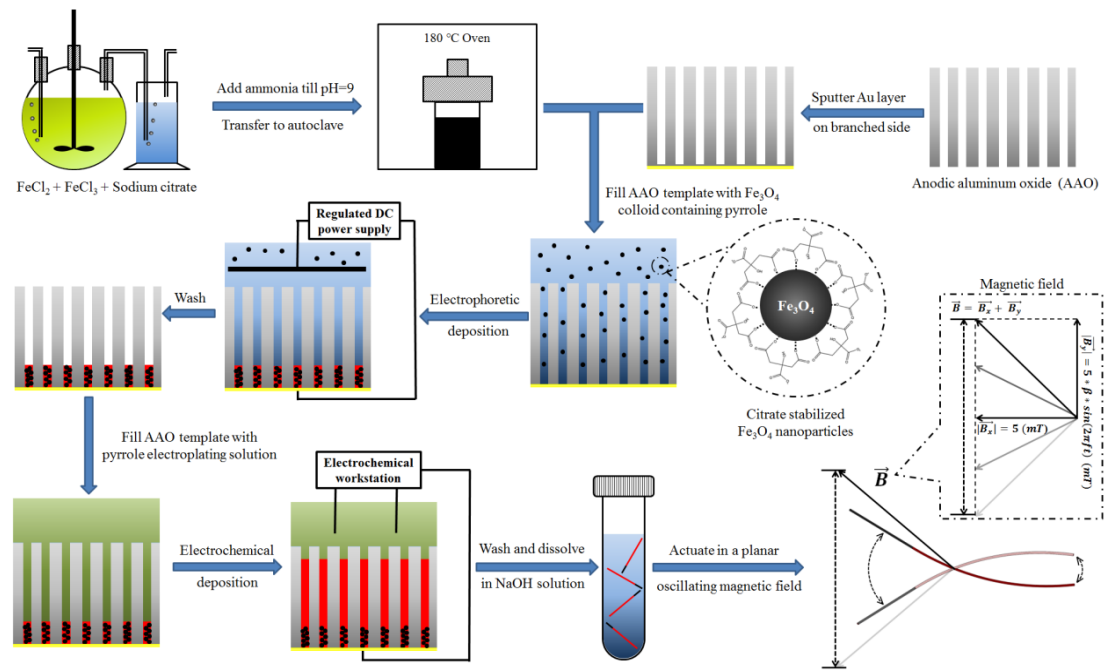

**Figure 2:** Schematic of the fabrication process of the microswimmer and the input planar oscillating magnetic field

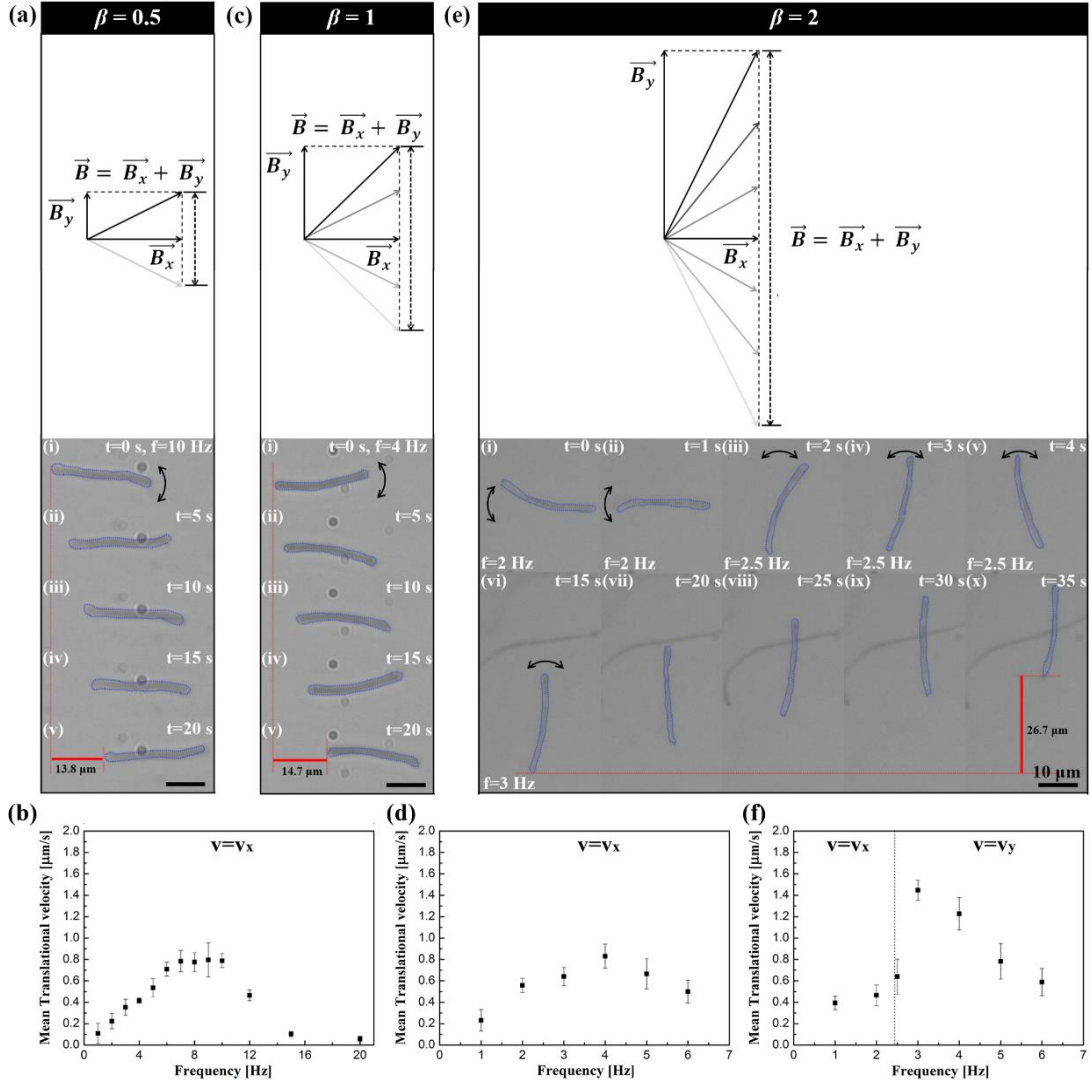

**Figure 3:** The microswimmer's motion status under an oscillating magnetic field with different values of *β* and *f*. (a) Motion snapshots of the microswimmer for *β*=0.5, *f*=10 Hz. (b) Average speed of the microswimmer along *x* direction as a function of frequency *f* for *β*=0.5. (c) Motion snapshots for *β*=1, *f*=4 Hz. (d) Average speed of the microswimmer along *x* direction as a function of frequency *f* for *β*=1. (e) (i)-(v) display the 90°-transition in oscillating direction of microswimmer for *β*=2 while *f* is increased from 2 to 2.5 Hz. (vi)-(x) show the dynamic motion of microswimmer in *y* direction when *β*=2 and *f*=3 Hz. (f) Average speed in *x* and *y* directions as a function of frequency for *β*=2. In (a)(c)(e), the black arrows represent the oscillating direction of the microswimmer and all the scale bars are 10 μm. Full movies appear in the Supplementary Information (Movie S1, 2, 3). In (b)(d)(f), "*V*=$V_x$" or "*V*=$V_y$" represent that the microswimmer swims along the *x* or *y* direction, respectively. The average speeds and standard deviation values are obtained from at least three speeds obtained from an image sequence.

No significant difference compared to the condition when $\beta$=0.5 is found except for the maximal speed and optimal frequency values. At this time, a maximal speed of 0.75-0.96 μm/s is obtained when the frequency of the magnetic field is 4 Hz. Remarkably, when we further increase $\beta$ to 2, a difference in the motion of the microswimmer is observed. As depicted in the images of (i) and (ii) of Fig. 2(e), the microswimmer still oscillates about the *x* axis when *f* is lower than 2 Hz. After a time of 20 seconds, when *f* is slightly increased to 2.5 Hz, the microswimmer almost immediately changes its oscillating direction from the *x* axis to the *y* axis, which is shown in (iii)-(v) of Fig. 2(e). This new motion condition (oscillations about the *y* axis) is very stable and further increasing *f* does not have any influence on it. Moreover, except for oscillating direction, 90° transition in the swimming direction of microswimmer also occurred. Images of (vi)-(x) of Fig. 2(e) show that the microswimmer starts swimming along the *y* direction with a velocity of ~1.34 μm/s (at *f*=3 Hz), which is comparable (and even larger) to the translational speed in *x* direction shown in Fig. 2(a) and (c). Finally, Fig. 2(f) shows a plot of the mean swimming speed as a function of frequency. We conclude that the 90°-transition in both oscillating orientation and swimming direction of the microswimmer can be obtained for a fixed value of $\beta$ by increasing the frequency *f* above a critical value. This finding is in agreement with our numerical simulations above. Next, we conduct a series of experiments in order to find the dependence of the transition conditions on both $\beta$ and *f*. Under different values of $\beta$ (from 0.5 to 3.5), the critical values of frequency, above which the microswimmer exhibits transition of motion to *y* direction, are shown in Fig. 3(a). In our experiments, when $\beta$ is lower than 2 the transition does not occur for any frequency, as indicated by the empty left region in Fig. 3(a). It is found that the critical frequency decreases with increasing $\beta$. For better illustration, the transition of oscillating orientation under varying frequency for $\beta$=3 is shown in Fig. S4 and Movie S4 in the SI. When the frequency *f* is held constant while $\beta$ is changed incrementally, we observe that an optimal value of $\beta$ exists, for which maximal swimming speed is achieved. This optimal speed is found to be within the region of "$V=V_y$", where the microswimmer swims along the *y* direction. For example, the swimming speed as a function of $\beta$ for *f*=3 Hz is shown in Fig. 3(b). For 0<$\beta$<2, increasing $\beta$ leads to the increase of velocity in the *x* direction. For $\beta$>2, the microswimmer begins to move in the *y* direction, and it is found that microswimmer exhibits a maximal speed of 1.52-1.69 μm/s when β=2.5. The dynamic motion with those actuation parameters are also shown in Fig. S5 and Movie S5 in the Supplementary Information. The existence of this optimal value of $\beta$, which has not been observed in previous works[16,17], is also in agreement with our numerical simulations.

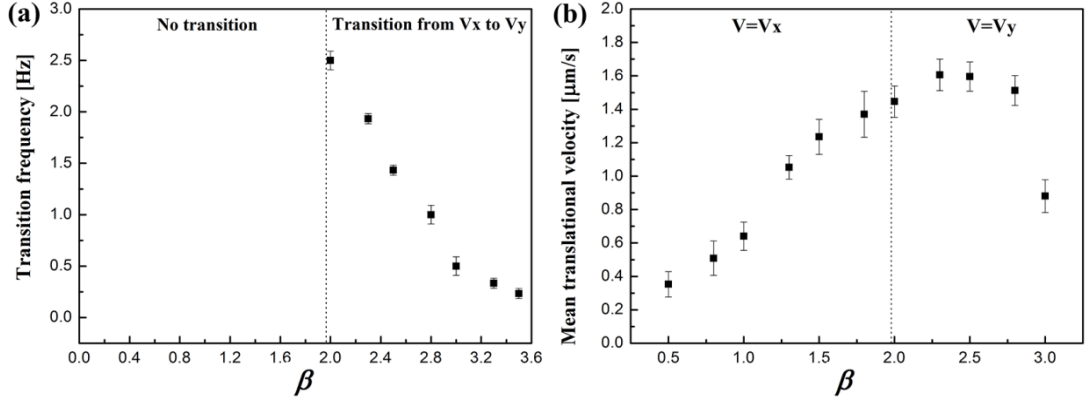

**Figure 4:** (a) The critical transition frequency $f$ as a function of the amplitude ratio $\beta$. Standard deviation values are obtained by repeating the 90°-transition in oscillating orientation three times. (b) The swimming speed of the microswimmer in $x$ and $y$ directions as a function of $\beta$ for $f=3$ Hz. The average speeds and standard deviation values are obtained from at least three speeds calculated from an image sequence.

**Theoretical two-link model**

In order to provide a theoretical explanation to the new findings observed in this microswimmer, we study the simplest possible model that describes our swimmer - two slender rigid links connected by an elastic joint, which move in ($x,y$) plane only (Fig. 5a). The length of each link is $l$, and their cross-section radius is $a$. A similar model has been studied previously for a microswimmer with *ferromagnetic* links[22], and it has been shown that the external magnetic actuation enables this swimmer to overcome the well-known *scallop theorem*[8] and generate non-reversible undulations that result in net swimming motion. In our work, the rightmost link (head) is made of a *superparamgnetic* material with anisotropy of $\Delta\chi$ and director $\hat{\mathbf{t}}$ which is aligned with the link's longitudinal axis. The magnetic field $\mathbf{B}(t)$ polarizes the link and induces a magnetic moment which, in turn, interacts with the field and generates a torque $\mathbf{L}$ given by $\mathbf{L} = \Delta\chi v (\hat{\mathbf{t}} \times \mathbf{B})(\hat{\mathbf{t}} \cdot \mathbf{B})$, where $v$ is the link's volume. Since the magnetic field $\mathbf{B}$ also lies in ($x,y$) plane, the torque it generates on the link simplifies to (see SI):

$$\mathbf{L} = \frac{1}{2}\Delta\chi v B^2 \sin(2\gamma)\hat{\mathbf{z}} \qquad (2)$$

where $B$ is the magnitude of the field $\mathbf{B}$ and $\gamma$ is the relative angle between its direction and the link's axis $\hat{\mathbf{t}}$ (Fig. 5a). This implies that the magnetic torque $\mathbf{L}$ vanishes when the directions of the magnetic field and the link are either aligned ($\gamma=\{0,\pi\}$) or perpendicular ($\gamma=\pm\pi/2$). It can readily be seen that for a *constant* magnetic field $\mathbf{B}$, the aligned orientations are stable while the perpendicular ones are unstable. However, when $\mathbf{B}(t)$ is oscillating as in (1), interaction with effects of swimmer's elasticity and fluid's viscosity changes the stability characteristic in a complicated way, as explained below. The microswimmer's dynamic equations of motion can be explicitly obtained as a system of nonlinear first-order differential equations (see SI for detailed derivation). The system's dynamics can be analyzed asymptotically in different physical limits, as follows. In the limit of small amplitude ratio $\beta$ of the magnetic field in (1), the system's solution can be expanded as power series in $\beta$ using perturbation expansion[23] (see SI). This solution involves stable oscillations about $\theta=\phi=0$, and swimming in $x$ direction, whereas oscillations about $\theta=90°$ are unstable. Leading-order expression for the mean swimming speed in $x$ direction is obtained as:

$$V_x = \beta^2 \frac{t_k \omega^2 l}{\omega^2 \left(t_k^2 \left(64 t_m^2 \omega^2 + 25\right) + 64 t_k t_m + 64 t_m^2\right) + 1} + O(\beta^4) \tag{3}$$

where $t_m, t_k$ are characteristic time scales of the system, given as:

$$t_m = \frac{c_t l^2}{6\Delta\chi v B_x^2} = \frac{viscous}{magnetic}, \quad t_k = \frac{c_t l^2}{12k} = \frac{viscous}{elastic} \quad \text{where } c_t = \frac{2\pi\mu l}{\log(l/a)} \tag{4}$$

and $\mu$ is the fluid's dynamic viscosity. Eq. (3) captures the resonance-like frequency dependence of the swimmer for low values of $\beta$, similar to that found in our simulations and experiments, and in agreement with Dreyfus et al[10]. Next, we study the transitions that occur when $\beta > 1$, where this small-amplitude approximation is no longer valid. We first consider the limit of very fast oscillations of the magnetic field, i.e. $\omega^{-1} \ll t_m, t_k$. Using the method of multiple scales[23], we define fast and slow nondimensional time scales $t_f = \omega t$ and $t_s = t/t_m$ and separate the system's dynamics into fast oscillations superimposed on a slowly evolving average solution. It can then be shown (see SI) that the slow-time average dynamics of the two angles $\bar{\theta}(t_s), \bar{\phi}(t_s)$ can be obtained asymptotically as:

$$\frac{d\bar{\theta}}{dt_s} = \frac{4\alpha\bar{\phi}(\cos(\bar{\phi})+3)^2 - (\beta^2-2)\sin(2\bar{\theta})(\cos(2\bar{\phi})+19)}{8(\cos(2\bar{\phi})-17)}$$

$$\frac{d\bar{\phi}}{dt_s} = \frac{(\cos(\bar{\phi})+3)\left(4\alpha\bar{\phi} - (\beta^2-2)\sin(2\bar{\theta})\right)}{8(\cos(\bar{\phi})-3)} \quad \text{where } \alpha = t_m/t_k. \tag{5}$$

The system of slow dynamics in (5) has only two equilibrium states, $\{\bar{\theta}_e = \bar{\phi}_e = 0\}$ and $\{\bar{\theta}_e = 90°, \bar{\phi}_e = 0\}$, which correspond to oscillations about $x$ and $y$ directions, respectively. Moreover, linearization of (5) about the equilibrium points reveals that $\bar{\theta}_e = 0$ is a stable equilibrium for $\beta < \sqrt{2}$, whereas $\bar{\theta}_e = 90°$ becomes stable for $\beta > \sqrt{2}$ (see SI). This is in agreement with the results obtained in Roper et al[16]. Furthermore, averaging over the fast oscillations gives an approximation of the mean swimming speed of the swimmer in $x$ and $y$ directions as (see SI):

$$V_x = 2\pi l f \left(\cos(\bar{\theta}_e) + \cos(3\bar{\theta}_e)\right) \left(\frac{1}{128}\alpha\beta^2\epsilon^3 + \frac{\alpha\beta^2\epsilon^5\left(-9(77-384\alpha)\beta^2 - 72(64\alpha(\alpha+1)+25) - 625\beta^4\right)}{589824}\right) \tag{6}$$

$$V_y = 2\pi l f \left(\sin(\bar{\theta}_e) - \sin(3\bar{\theta}_e)\right) \left(\frac{1}{128}\alpha\beta^2\epsilon^3 + \frac{\alpha\beta^2\epsilon^5\left(-9(384\alpha+77)\beta^2 - 72(64(\alpha-1)\alpha+25) - 625\beta^4\right)}{589824}\right)$$

where $\epsilon = (\omega t_m)^{-1}$. This confirms that oscillations about $\bar{\theta}_e = 0$ correspond to motion in $x$ direction, whereas oscillations about $\bar{\theta}_e = 90°$ correspond to motion in $y$ direction. Importantly, (6) also indicates that there exist an optimal value of $\beta$ that maximizes the mean speed, as corroborated in our experimental observations. However, this limit of fast oscillations does not capture the dependence of the critical value of value of $\beta$ on the frequency $f$.

In case where the oscillation frequency $f$ is not so large, the system's equations can be simplified by assuming a limit of stiff spring, $t_k \ll t_m$, which implies small joint angle $\phi(t)$. Linearizing by $\phi$, one obtains a scalar second-order differential equation for $\theta(t)$ as (see SI):

$$32\ddot{\theta} + \left(32\alpha + 10\cos(2\theta)\left(\beta^2\cos(2t\omega) - \beta^2 + 2\right) + 40\beta\sin(2\theta)\sin(t\omega)\right)\dot{\theta}$$
$$+ \left(\alpha(2-\beta^2) + \alpha\beta^2\cos(2t\omega) - 10\omega\beta^2\sin(2t\omega)\right)\sin(2\theta)$$
$$= 4\left(5\beta\omega\cos(t\omega) + \alpha\beta\sin(t\omega)\right)\cos(2\theta)$$

(7)

Equation (7) is a second-order system with parametric excitation[23], whose structure is remarkably similar to the well-known equation of *Kapitza pendulum* with tilted base oscillations[19,24], see Figure 6a. Using numerical integration of (7), periodic solutions can be computed, and their orbital stability properties can be obtained by evaluating the Floquet multipliers of the linearized variational equation for small perturbations about each periodic solution (see SI). This enables plotting stability regions of the two different solutions in the plane of ($\beta$,$f$), as shown in the stability transition curves in Figure 6b. These curves show that the stability transitions depend on both amplitude $\beta$ and frequency $f$ of the oscillations (in analogy to Kaptiza pendulum). It can be seen that for large frequency $f$, the transitions value approaches $\beta \to \sqrt{2}$, in agreement with the results of Roper[17] and our analysis. Another interesting observation is the existence of a significant region of *bistability*, where the solutions of swimming in *x* and *y* directions are both locally stable, with different regions of attractions in terms of initial conditions, as demonstrated in Figure 5d. Nevertheless, it is often practically difficult to observe co-existence of solutions in a real microswimmer. This is because one cannot precisely enforce any desired initial conditions, while some periodic solutions may have very narrow regions of attraction.

Calibrating parameters of the theoretical models by fitting to the experiments gives $t_m = t_k = 0.1 sec$, and $l = 25 \mu m$. Figure 6c plots a comparison of the stability transition curves obtained experimentally (squares) with our theoretical analysis, where the solid curves denote the theoretical transition curves of local stability, and the region enclosed between them is a region of theoretical bistability. In addition, the dashed line plotted in Fig. 6c is the curve of transition between swimming in *x* and *y* directions, under specific initial conditions of θ(0)=ϕ(0)=0. That is, the swimmer is initially straightened and aligned with *x* axis. This initial condition is practically realized in our experiments by initializing the microswimmer under a constant magnetic field in *x* direction. Therefore, this curve gives a meaningful comparison to our experiment, showing a remarkable agreement with the theoretical two-link model. Finally, Fig. 6d shows a comparison of the net swimming speeds $V_x$ and $V_y$ obtained from the two-link model (solid lines) with our experimental results, showing an excellent agreement.

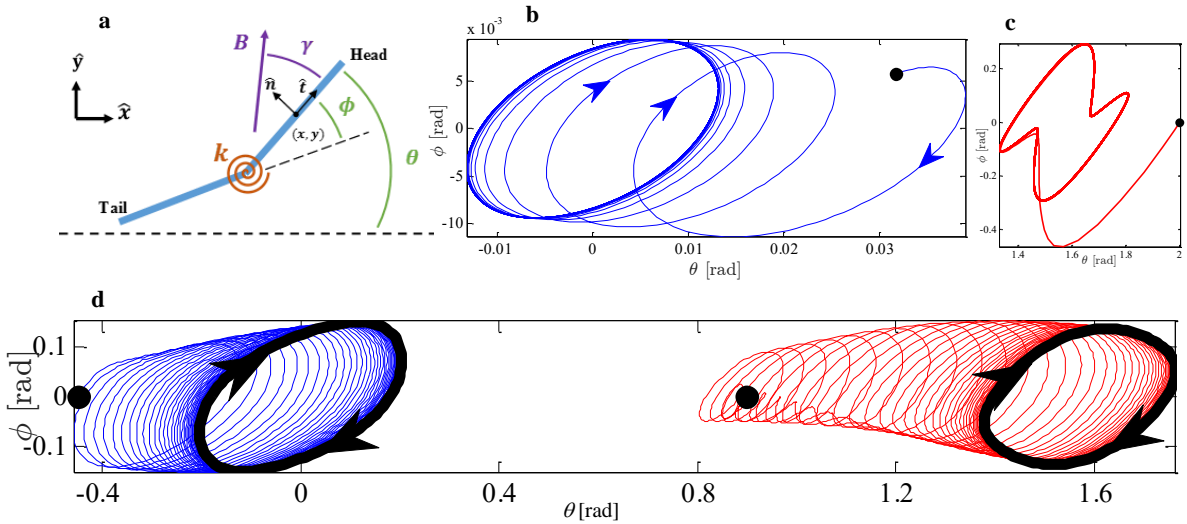

**Figure 5:** (a) The planar two-link microswimmer model. The head link is superparamagnetic. (b-d) Solution trajectories of the two-link microswimmer in the plane of angles ($\theta, \phi$), for model parameters $t_k=0.1$, $t_m=1$, and $f=0.16$. (b) Solutions under small oscillations $\beta=0.1$ - convergence to a periodic trajectory where both $\theta(t)$ and $\phi(t)$ oscillate about $0°$ and the swimmer moves in *x* direction. (c) Solutions under large oscillations $\beta=10$ - convergence to a periodic trajectory where $\theta(t)$ oscillates about $90°$ while $\phi(t)$ oscillates about $0°$, and the swimmer moves in *y* direction. (d) Solutions under intermediate oscillations $\beta=1.45$ - bistability conditions where the periodic solutions around $\theta=0°$ and $\theta=90°$ are both stable, and convergence is determined by initial conditions.

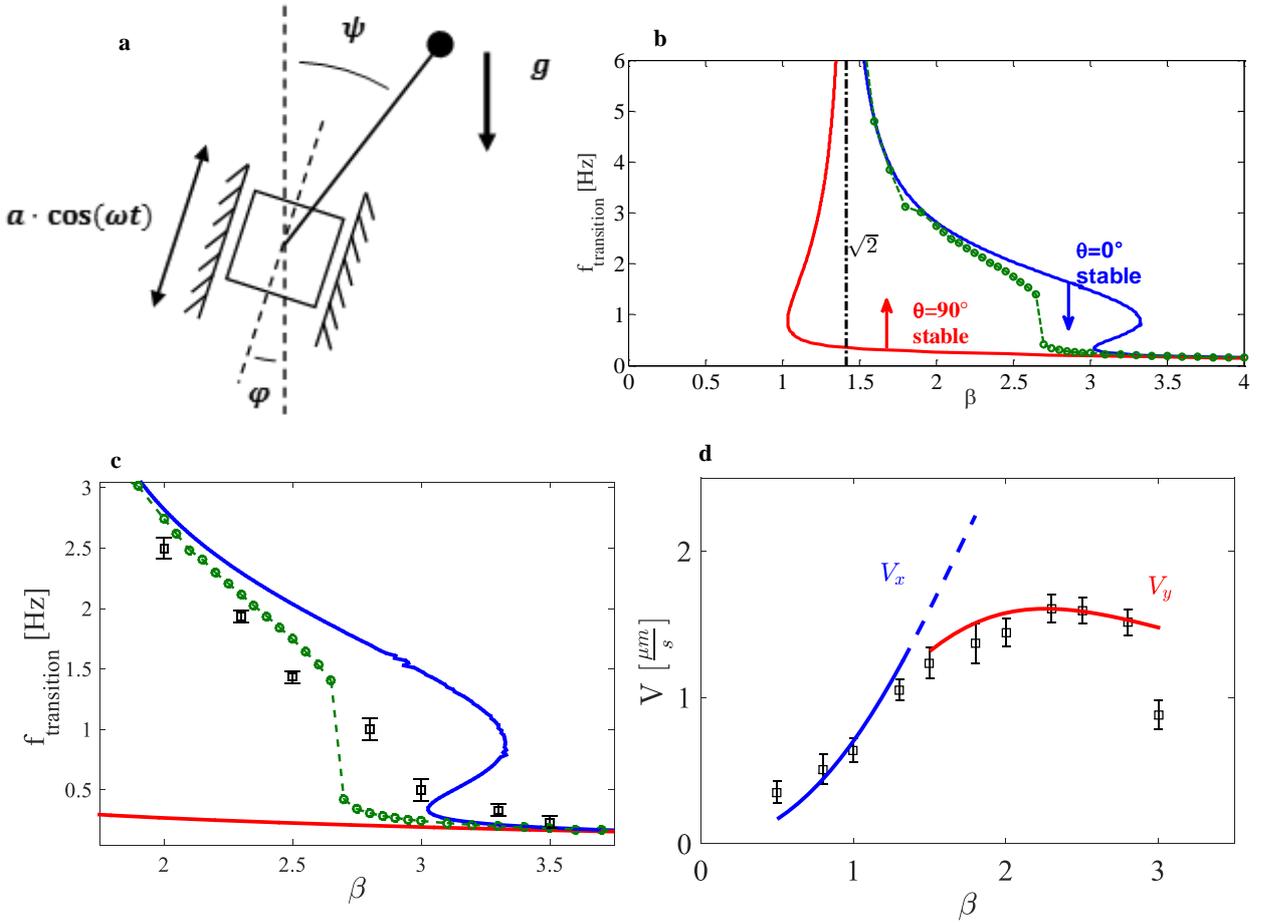

**Figure 6:** (a) The example of Kapitza pendulum[19,24], an inverted pendulum with an oscillating pivot. The pendulum has mass *m* and length *l*. Its tilt angle is denoted by *ψ* and gravity acceleration is *g*. The pivot of the pendulum is oscillating along a direction with inclination angle *φ*, and displacement of *a · cos(ωt)*. The equation of motion of this pendulum is given by $\ddot{\psi} - \left(\frac{g}{l} + \frac{a\omega^2}{l}\cos\varphi \cos(\omega t)\right)\sin\psi = -\frac{a\omega^2}{l}\sin\varphi \cos(\omega t) \cos\psi$. The structure of this equation is remarkably similar to eq. (7), where the pendulum angle *ψ* is replaced by *2θ*. Nevertheless, equation (7) also contains an additional "parametric damping term", i.e. proportional to $\dot{\theta}$. (b) Plot of stability transition curves in the plane of parameters *β* and *f*, the amplitude and frequency of the magnetic field's oscillations. The lower curve is stability limit of swimmer's oscillations about θ=0°, and the upper curve is stability limit of swimmer's oscillations about θ=90°. This plot shows that stability transitions depend on both *β* and *f*, as observed in our experiments. The region between the two curves is *bistability* regime, where both solutions are stable under different initial conditions. The dashed curve represents the transition curevs under practical initial conditions of θ=ϕ=0. (c) Zoom into the plot of stability transition curves, compared to our experimental results. (d) Plot of net swimming speed as a function of *β* for *f*=3Hz, compared to our experimental results. The model correctly captures the maximum in $V_y$ for optimal value of *β*. The dashed line is the bistable solution of $V_x$ which has not been reached under practical initial conditions of θ=ϕ=0.

In summary, we have studied stability transitions in microswimmers with superparamagnetic links under planar oscillations of an external magnetic field. Unlike previous studies, we have shown that the conditions for transition depend on various parameters, including amplitude and frequency of the oscillations. We used numerical simulations and experimental measurements of a microswimmer with superparamagnetic head. A simple two-link model enabled asymptotic analysis of the swimmer's nonlinear dynamics, and provided an explanation to the stability transitions which are induced by a nonlinear parametric excitation.

**Acknowledgment**
The work of D.J. and L.Z. was supported by the Early Career Scheme (ECS) grant with Project No. 439113 and the General Research Fund (GRF) with Project No. 14209514 and 14203715 from the Research Grants Council (RGC) of Hong Kong SAR. The work of Y.H. and Y.O. was supported by the Israeli Science Foundation (ISF) under grant no. 567/14.

**Author contributions**
Y.O. and L.Z. initiated the research. Y.O. and Y.H. performed the numerical simulations. L.Z. and D.J. designed and fabricated the microswimmer. L.Z. and D.J. designed and provided the experimental magnetic setup for manipulation. D.J. conducted the magnetic actuation experiments and performed the analysis of the dynamic motion of microswimmers. Y.O. and Y.H. proposed the two-link model and achieved the theoretical analysis. D.J., Y.H., L.Z. and Y.O. cowrote the manuscript. Y.O. and L.Z. supervised the work and gave critical input. All authors contributed to discussions.

**Additional information**
Supplementary information is available in the online version of the paper. Reprints and permissions information is available online at www.nature.com/reprints.
Correspondence and requests for materials should be addressed to Y.O. or L.Z.

**Competing financial interests**
The authors declare no competing financial interests.

# Supplementary Information for manuscript: "Nonlinear parametric excitation effect induces stability transitions in swimming direction of flexible superparamagnetic microswimmers"

Yuval Harduf, Dongdong Jin, Yizhar Or and Li Zhang

# Part 1 - Experiments

This part contains detailed information on the experiments. In particular, it presents the fabrication process and characterization of the flexible superparamagnetic microswimmers, as well as the magnetic actuation experiments.

**1.1 Chemicals**

The chemicals of iron(III) chloride hexahydrate (99%), iron(II) chloride tetrahydrate (99%), sodium citrate dihydrate (99%), ammonia solution (25-28%), citric acid monohydrate (99.5%), pyrrole (99%) and sodium hydroxide (97%) were purchased from Aladdin Chemicals. Anodic aluminum oxide (AAO) template with average pore diameter of ~200 nm and thickness of 60 μm was purchased from Shanghai Shangmu Technology. All the chemicals were used without further purification.

**1.2 Synthesis and characterizations of $Fe_3O_4$ nanoparticles**

The $Fe_3O_4$ nanoparticles were prepared by hydrothermal method. In a typical procedure, 30 mL of ultrapure water (18.2 MΩ*cm) was first purged with dry argon gas in a round-bottom flask while stirring for 30 min to remove the oxygen in water. Subsequently, 0.5 ml of concentrated hydrochloric acid, 30 mmol of $FeCl_3$ $6H_2O$, 15 mmol of $FeCl_2$ $4H_2O$ and 30 mmol of sodium citrate dihydrate were added into the flask. After the chemicals were completely dissolved, the pH value of obtained solution was adjusted as 9 by gradually adding ammonia, affording a black precipitate. The black product was poured into a teflon-lined stainless steel autoclave which was then sealed and heated at 180 °C for 2 h. Finally, the nanoparticle product was collected by centrifugation at 10000 rpm for 5 min and washed with ultrapure water for three times, followed by redispersing in ultrapure water prior to use.

Before fabricating microswimmers, the morphology, size distribution, magnetic property and surface charge distribution of the as-prepared $Fe_3O_4$ nanoparticles were investigated (shown in Supplementary Fig. 1(a-d)) using transmission electron microscope (Tecnai Spirit 12, FEI), Zeta Sizer (Nano ZS, Malvern), and vibrating sample magnetometer (VSM 7410, Lake Shore).

Typical TEM micrograph of the $Fe_3O_4$ nanoparticles shown in Supplementary Fig. 1a displayed that $Fe_3O_4$ nanoparticles had an average diameter of 10-20 nm. When we dispersed $Fe_3O_4$ nanoparticles in water, however, aggregation of $Fe_3O_4$ nanoparticles would occur, leading to the increasing size of $Fe_3O_4$ particles. It could be observed from Supplementary Fig. 1b that the actual diameter of most $Fe_3O_4$ nanoparticles (~68.4%) in electrophoresis bath was between 40 and 100 nm, and the average size of overall $Fe_3O_4$ nanoparticles was calculated as ~80 nm. Therefore, given that the average pore size of AAO template was ~200 nm, the possibility of assembling the as-prepared $Fe_3O_4$ particles into AAO template might be realized. Supplementary Fig. 1c showed the magnetic hysteresis loop of the $Fe_3O_4$ nanoparticles at room temperature. The $Fe_3O_4$ nanoparticles were superparamagnetic with a saturation value of 65.4 emu/g. Moreover, the average zeta potential of $Fe_3O_4$ nanoparticles was calculated as ~-50 mV from Supplementary Fig. 1d, demonstrating that $Fe_3O_4$ nanoparticles could be electrophoretic deposited in working electrode by applying a positive voltage between working electrode and counter electrode.

### 1.3 Synthesis of microswimmers

Microswimmer with a superparamagnetic head was synthesized by electrophoretic codeposition of $Fe_3O_4$ nanoparticles and polypyrrole followed by electrochemical deposition of polypyrrole. A thin gold film (~100 nm) was first coated on one side of AAO templet using magnetron sputtering and the AAO template was then assembled in a homemade teflon plating cell (shown in Supplementary Fig. 2a) to serve as a working electrode. Subsequently, the plating cell was put in a modified vacuum drier as shown in Supplementary Fig. 2b. For the electrophoretic deposition of $Fe_3O_4$ nanoparticles based superparamagnetic head, 10 mL of aqueous solution containing ~1 mg of $Fe_3O_4$ nanoparticles and 1 mmol of pyrrole acting as electrophoresis bath was added into the plating cell under vacuum to facilitate the nanoparticles to enter into the pores of AAO template. A thin stainless steel sheet was employed as a counter electrode, and the distance between working electrode and counter electrode was 10 mm. Then the electrophoretic deposition was performed at a constant electric potential of +50 V for 3 h. After the reaction was completed, the plating solution was poured and the assembled AAO template was washed by being immersed in ultrapure water for 30 min. Next, for the electrochemical deposition of polypyrrole nanowire based flexible tail, a Pt wire and an Ag/AgCl electrode were used as counter and reference electrodes, respectively. Then polypyrrole nanowires were electropolymerized at +0.80 V for 3 C from a plating solution containing 0.1 M citric acid and 0.1 M pyrrole. Finally, the AAO template was taken out from plating cell and dissolved in 3M NaOH for 30 min to obtain the microswimmers, which was then collected by centrifugation at 10000 rpm for 3 min and washed with ultrapure water for three times. All microswimmers were stored in ultrapure water at room temperature when not in use. Typical scaning electron microscopy (SM-7800F, JEOL) images of as-prepared microswimmers were shown in Supplementary Fig. 1e, f.

**1.4 Magnetic actuation experiment**

The magnetic control experiments were conducted using a 3-axis Helmholtz electromagnetic coil setup as shown in Supplementary Fig. 3. The control signals were generated by a Sensoray 826 card and then the current was amplified to generate magnetic fields in the coils. There were three pairs of coils in the setup and in our experiments, the current through the middle coils was constant to generate a constant x vector, the current through the big coils was alternating to generate an oscillating y vector, while the current through the small coils was 0. In this way, a planar oscillating magnetic field could be generated.

Before the swimming experiments, microswimmer was put in an open tank filled with ultrapure water by a pipette which was then put in the centre of the coil pairs. During the actuation experiments, a constant magnetic field of 5 mT in x direction ($\vec{B} = \vec{B_x}$) was first set for 5s acting as the initial condition. Then an oscillating y vector with a certain value of $\beta$ and 0.1 Hz of $f$ was generated together with the constant x vector. We gradually increased the frequency and tried to look for the 90° transition in the oscillating orientation and swimming direction of microswimmer and calculate the swimming velocity. Subsequently, we turned off the magnetic field, set a constant x vector for 5s renewedly, and then changed the value of $\beta$, and conducted the magnetic actuatuion experiments in the same way. Each status with a certain value of $\beta$ and $f$ was kept for at least 5s.

# Part 2 - Theory

This part contains detailed information on the theoretical analysis. In particular, it presents the multi-link model of the microswimmer and its numerical simulations, as well as the simplified two-link model and the asymptotic analysis of its dynamics.

**2.1 Discrete multi-link model**

Our model of the swimmer as presented in Supplementary Fig. 6 consists of a chain of $n$ slender links ("tail"), attached to a non-magnetic spherical "head", which represents the load. The links are connected by rotary joints with linear torsion springs, and carry superparamagnetic beads. The swimmer is submerged in a viscous fluid and is actuated by an external magnetic field. While Dreyfus's model[1] considered the tail as a continuous filament of length $L$, we treat it here as a chain of discrete, rigid links, with equal lengths $l = \frac{L}{n}$ and a circular cross-section of radius $a$. In both cases the red blood cell is modeled as a rigid sphere of radius $r_h$.

We define a world-fixed reference frame whose axes are $\hat{\mathbf{x}}, \hat{\mathbf{y}}, \hat{\mathbf{z}}$. Only planar motion of the swimmer in $\hat{\mathbf{x}} - \hat{\mathbf{y}}$ plane is considered, and all rotations are about $\hat{\mathbf{z}}$ axis. We attach a reference frame to each link, whose axes are $\hat{\mathbf{t}}_i, \hat{\mathbf{n}}_i, \hat{\mathbf{z}}$, where the direction $\hat{\mathbf{t}}_i$ is aligned with the longitudinal axis of the link and $\hat{\mathbf{n}}_i$ is perpendicular to it. The location of each link is defined as its geometrical center and its orientation is defined as the angle between its axis $\hat{\mathbf{t}}_i$ and the $\hat{\mathbf{x}}$ axis. We denote $(x \quad y)^T$ as the location of the head link in the world-fixed frame, $\theta$ is the angle between the head's axis $\hat{\mathbf{t}}_1$ and the $\hat{\mathbf{x}}$ axis, and $\phi_i$ is the relative angle between the axis of the $i^{th}$ link $\hat{\mathbf{t}}_i$ and the axis of the $i+1^{th}$ link $\hat{\mathbf{t}}_{i+1}$.

### 2.1.1 The magnetic torques acting on the links

The swimmer is actuated by a uniform, planar magnetic field denoted by $\mathbf{B}(t)$, which is assumed to vary within the $\hat{\mathbf{x}} - \hat{\mathbf{y}}$ plane and has the form:

$$\mathbf{B} = \begin{pmatrix} 1 \\ \beta \sin(\omega t) \end{pmatrix} B_x \tag{S1}$$

In Dreyfus et al[1] the paramagnetic beads have a preferred magnetization direction. In the assembly process, it is assumed that all of the beads' directions align with the backbone of the continuous filament. It is also assumed that each bead experiences magnetic interactions only with the two beads adjacent to it. The magnetic torque per unit length in Ref [1] is of the form:

$$\frac{d\mathbf{L_m}}{ds} = \frac{\pi r_b^2}{3\mu_0} \left( \frac{\chi_= - \chi_\perp + \chi_= \chi_\perp / 4}{(1 - \chi_= / 6)(1 + \chi_\perp / 12)} \right) \left( \hat{\mathbf{t}}(s) \times \mathbf{B} \right) \left( \hat{\mathbf{t}}(s) \cdot \mathbf{B} \right) \tag{S2}$$

Where $s$ is a coordinate along the filament's length, $\chi_=, \chi_\perp$ are the susceptibility components of a single bead in the preferred magnetization direction and in the direction perpendicular to it, respectively, $\mu_0$ is the vacuum permeability, $r_b$ is the beads' radius and $\hat{\mathbf{t}}(s)$ is the local unit vector tangent to the filament. In our model, where the continuous filament is regarded as a discrete chain of links, the torque acting upon each link is obtained by integrating (S2) over the length of a link, and has the form:

$$\mathbf{L_i} = \frac{\pi r_b^2}{3} l \tilde{\chi} \left( \hat{\mathbf{t}}_i \times \mathbf{B} \right) \left( \hat{\mathbf{t}}_i \cdot \mathbf{B} \right) \tag{S3}$$

Where:

$$\tilde{\chi} = \frac{1}{\mu_0} \left( \frac{\chi_= - \chi_\perp + \chi_= \chi_\perp / 4}{(1 - \chi_= / 6)(1 + \chi_\perp / 12)} \right) \tag{S4}$$

### 2.1.2 Internal torques acting at the joints

We denote the torques acting at the joints as $\tau_i$. Since the links are connected by torsion springs with a linear spring constant $k$, the torques acting at the joints are given by:

$$\tau_i = -k\phi_i \tag{S5}$$

We would like to determine the springs' constant in such a way that the deflection of our "discrete filament" will be comparable with the deflection of Dreyfus' filament in Ref [1]. We use Euler-Bernoulli beam theory[2] to model the deflection of Dreyfus' filament, and we assume that its properties are constant along its length. Looking at an infinitesimal cross-section of the beam we get:

$$\kappa \frac{d\theta(s)}{ds} = M(s) \tag{S6}$$

Where $\kappa$ is the filament's bending rigidity from Dreyfus's model. We assume that the bending torque is constant along the beam and integrate along the beam's length:

$$\int_0^l \kappa \frac{d\theta}{ds} ds = \int_0^l M ds \rightarrow \kappa\theta = Ml \tag{S7}$$

And so we obtain our spring constant to be:

$$k = \frac{M}{\theta} = \frac{\kappa}{l} = \frac{\kappa n}{L} \tag{S8}$$

**2.1.3 The hydrodynamic forces acting on the swimmer**

Due to the extremely small Reynolds number of the problem, the fluid is governed by Stokes flow, which in turn dictates that the drag forces are linear in the velocity[3]. We model the drag forces using resistive force theory (RFT)[4, 5] which gives the following relation:

$$\mathbf{f_i} = -c_t(\mathbf{u_i}\cdot\hat{\mathbf{t}_i})\hat{\mathbf{t}_i} - c_n(\mathbf{u_i}\cdot\hat{\mathbf{n}_i})\hat{\mathbf{n}_i}, \quad M_i = -c_m\omega_i \tag{S9}$$

Where $\mathbf{f_i}, M_i$ is the force vector and torque acting upon the $i^{th}$ link, $\mathbf{u_i}$ is the link's velocity vector, $\omega_i$ is the link's angular velocity. $c_t, c_n, c_m$ are the tangential, normal and rotational drag coefficients accordingly. According to resistive force theory[4, 5], for a slender link these drag coefficients are:

$$c_n \approx 2c_t \approx \frac{4\pi\mu l}{\log(l/a)}, \quad c_m = \frac{1}{6}c_t l^2 \tag{S10}$$

And for the spherical head these drag coefficients are[3]:

$$c_n = c_t = 6\pi\mu r, \quad c_m = 8\pi\mu r^3 \tag{S11}$$

Where $\mu$ is the fluid's viscosity. In Dreyfus' model[1] the drag is modeled in the same manner and the drag coefficients are the same, only defined per unit length. It is assumed that the hydrodynamic interaction between links is negligible and so the hydrodynamic forces on a single link are only a result of that link's velocity.

**2.1.4 Formulating the equations of motion**

We define a body coordinates vector $\mathbf{q_b} = (x \quad y \quad \theta)^T$ and a shape coordinates vector $\mathbf{q_s} = (\phi_1 \quad \phi_2 \quad \cdots \quad \phi_n)^T$. Both vectors compose the full vector of generalized coordinates $\mathbf{q} = (\mathbf{q_b}, \mathbf{q_s})$. We define the orientation of the $i^{th}$ link as the angle between the $\hat{\mathbf{t}}_i$ axis and the $\hat{\mathbf{x}}$ axis as $\alpha_i$ and write them in terms of the body and shape coordinates:

$$\alpha_i = \theta + \sum_{j=1}^{i-1} \phi_j \tag{S12}$$

We denote the location of the swimmer as $\mathbf{r_b}$, the location of each link as $\mathbf{r_i}$ and the location of each joint as $\mathbf{b_j}$. These vectors are given by:

$$\begin{aligned}
\mathbf{r_1} &= \mathbf{r_b} = \begin{pmatrix} x \\ y \end{pmatrix} \\
\mathbf{r_2} &= \mathbf{r_1} - \begin{pmatrix} r_h \cos(\alpha_1) + \dfrac{l}{2}\cos(\alpha_2) \\ r_h \sin(\alpha_1) + \dfrac{l}{2}\sin(\alpha_2) \end{pmatrix} \\
\mathbf{r_i} &= \mathbf{r_{i-1}} - \dfrac{l}{2}\begin{pmatrix} \cos(\alpha_{i-1}) + \cos(\alpha_i) \\ \sin(\alpha_{i-1}) + \sin(\alpha_i) \end{pmatrix}, \text{ for } i = 3 \ldots n
\end{aligned} \tag{S13}$$

$$\mathbf{b_j} = \mathbf{r_{j+1}} + \dfrac{l}{2}\begin{pmatrix} \cos(\alpha_{j+1}) \\ \sin(\alpha_{j+1}) \end{pmatrix} \tag{S14}$$

Next, we write the linear and angular velocities of each link in terms of the body and shape velocities, expressed in the body-fixed reference frame:

$$\mathbf{V_i} = \begin{pmatrix} \mathbf{u_i} \\ \omega_i \end{pmatrix} = \mathbf{T_i}(\phi)\dot{\mathbf{q}}_b + \mathbf{E_i}(\phi)\dot{\mathbf{q}}_s \tag{S15}$$

Where $\mathbf{u_i}$ is the vector of velocities and $\omega_i$ is the angular velocity of the $i^{th}$ link about the $\hat{\mathbf{z}}$ axis. The matrices $\mathbf{T_i}$ and $\mathbf{E_i}$ are given as follows[6]:

$$\mathbf{T_i} = \begin{pmatrix} \mathbf{I}_{2\times2} & -\mathbf{J}(\mathbf{r_i}-\mathbf{r_b}) \\ 0 \quad 0 & 1 \end{pmatrix}$$

$$\mathbf{E_i} = (\mathbf{E_{i1}} \quad \mathbf{E_{i2}} \quad \cdots \quad \mathbf{E_{im}}), \mathbf{E_{ij}} = I_{ij}\begin{pmatrix} \mathbf{J}(\mathbf{r_i}-\mathbf{b_j}) \\ 1 \end{pmatrix}$$

(S16)

Where $I_{ij} = \begin{cases} 1 \; for \; i > j \\ 0 \; for \; i \le j \end{cases}$ and $\mathbf{J} = \begin{pmatrix} 0 & -1 \\ 1 & 0 \end{pmatrix}$

We can now write relation (S9) in matrix form, using "resistance matrices":

$$\mathbf{F_{hyd,i}} = \begin{pmatrix} \mathbf{f_i} \\ M_i \end{pmatrix} = -\mathbf{R_i}(\phi)\mathbf{V_i} \tag{S17}$$

Where $\mathbf{R_i}$ are given explicitly as follows. For the spherical link (head):

$$\mathbf{R_1} = \pi\mu r_h \begin{pmatrix} 6 & 0 & 0 \\ 0 & 6 & 0 \\ 0 & 0 & 8r_h^2 \end{pmatrix} \tag{S18}$$

And for the slender links:

$$\mathbf{R_i} = \frac{2\pi\mu l}{\log(l/a)} \begin{pmatrix} 1+\sin(\alpha_i^2) & -\cos(\alpha_i)\sin(\alpha_i) & 0 \\ -\cos(\alpha_i)\sin(\alpha_i) & 1+\cos(\alpha_i^2) & 0 \\ 0 & 0 & \frac{1}{6}l_i^2 \end{pmatrix}, i>1 \tag{S19}$$

**2.1.5 Balance of forces and torques**

Since the swimmer is of micro-scale, the flow's Reynolds number is extremely small. Neglecting all inertial effects, the swimmer moves quasi-statically in equilibrium. Formulating a force and torque balance on the swimmer's body gives:

$$\sum \mathbf{T_i^T}(\mathbf{F_{hyd,i}} + \mathbf{F_{b_i}}) = 0 \tag{S20}$$

Substituting (S17) and (S15) into (S20) we get:

$$\mathbf{R}_{bb}\dot{\mathbf{q}}_\mathbf{b} + \mathbf{R}_{bu}\dot{\mathbf{q}}_\mathbf{s} = \sum \mathbf{T_i^T}\mathbf{F_{b_i}} \tag{S21}$$

Formulating torque balance at each joint leads to (see [6]):

$$\sum \mathbf{E_i^T}\left(\mathbf{F_{hyd,i}} + \mathbf{F_{b_i}}\right) + \boldsymbol{\tau}_s = 0 \tag{S22}$$

Where $\boldsymbol{\tau}_s = (\tau_1, \ldots \tau_n)$ satisfies $\boldsymbol{\tau}_s = -k\mathbf{q}_s$ from (S5). Substituting (S17), (S15) and (S5) into (S22) gives:

$$\mathrm{R}_{bu}^{\ T}\dot{\mathbf{q}}_\mathbf{b} + \mathrm{R}_{uu}\dot{\mathbf{q}}_\mathbf{s} = \sum \mathbf{E_i^T}\mathbf{F_{b_i}} - k\mathbf{q}_s \tag{S23}$$

Where:

$$\mathrm{R}_{bb} = \sum \mathbf{T_i^T}\mathrm{R_i}\mathbf{T_i}, \quad \mathrm{R}_{bu} = \sum \mathbf{T_i^T}\mathrm{R_i}\mathbf{E_i}, \quad \mathrm{R}_{uu} = \sum \mathbf{E_i^T}\mathrm{R_i}\mathbf{E_i}. \tag{S24}$$

It can be shown (see [6]) that the matrices $\mathbf{T_i}, \mathbf{E_i}$ in (S20) and (S22) are the same as the ones in equation (S15). Collecting the equations into matrix form, one obtains:

$$\mathbf{A(q)\dot{q}} = \mathbf{b(q,}t) \tag{S25}$$

where 
$$\mathbf{A(q)} = \begin{pmatrix} \mathrm{R}_{bb} & \mathrm{R}_{bu} \\ \mathrm{R}_{bu}^{\ T} & \mathrm{R}_{uu} \end{pmatrix}, \quad \mathbf{b(q,}t) = \begin{pmatrix} \sum \mathbf{T_i^T}\mathbf{F_{b_i}} \\ \sum \mathbf{E_i^T}\mathbf{F_{b_i}} - k\mathbf{q}_s \end{pmatrix} \tag{S26}$$

Equation (S25) is a system of $n+3$ first-order, time-dependent, nonlinear ordinary differential equations, which are linearly coupled. The solution of this system under given initial conditions can be obtained computationally via numerical integration.

**2.1.6 Numerical methods**

All of our numerical analyses were conducted using MATLAB's integration function ODE15s. In order to obtain the mean speed of the swimmer in a period we used the following scheme:

1. Set physical parameters

2. Integrate over $n_p$ periods of the magnetic field (typically $n_p$ is in the range of $100 \sim 1000$, in order for the system to reach steady state)

3. The mean swimming speed is calculated using the following formula:

$$\begin{aligned} V_x &= \frac{\omega}{2\pi}\left(x_{n_p} - x_{n_p-1}\right) \\ V_y &= \frac{\omega}{2\pi}\left(y_{n_p} - y_{n_p-1}\right) \end{aligned} \tag{S27}$$

In all of the simulations we used a swimmer whose tail consists of $n = 10$ links, and unless stated otherwise, the numerical values for the physical parameters used in the simulations are taken from the work of Dreyfus et al[1]. For the simulation results in Fig. 1b in the manuscript, these values are summarized in Supplementary Table 1.

For the simulation results in Fig. 1c in the manuscript, we used the "green squares" data from Supplementary Table 2, without a spherical head, under varying values of the magnetic field. For the simulation results in Fig. 1d and 1e, we used the "green squares" data from Supplementary Table 3, including a spherical head, under varying values of the magnetic field.

**2.2 The two-link model**

We now introduce the two-link model, which is the simplest possible model that captures the microswimmer's dynamics and stability transitions. It consists of two slender links connected by a single rotary joint with a torsion spring, as shown in Supplementary Fig.7. The microswimmer's "head" link is made of superparamagnetic material having uniaxial anisotropy with an easy axis along the longitudinal axis of the link $\hat{t}$. The susceptibility tensor of the uniaxial anisotropic link is:

$$\boldsymbol{\chi} = \chi_0 \mathbf{I} + \Delta\chi \left( \hat{\mathbf{t}}^T \hat{\mathbf{t}} - \frac{1}{3}\mathbf{I} \right) \tag{S28}$$

The magnetic field **B** polarizes the head link and induces a magnetic moment:

$$\mathbf{M} = \boldsymbol{\chi} \cdot \mathbf{B} v \tag{S29}$$

Where $v$ is the link's volume. The magnetic field generates a torque acting on the polarized link:

$$\mathbf{L} = \mathbf{M} \times \mathbf{B} = \Delta\chi v (\hat{\mathbf{t}} \times \mathbf{B})(\hat{\mathbf{t}} \cdot \mathbf{B}) = L_m \hat{\mathbf{z}} = \tfrac{1}{2} \Delta\chi v B^2 \sin(2\gamma) \hat{\mathbf{z}} \tag{S30}$$

Where $\gamma$ is the angle between the magnetic field's vector **B** and the link's direction $\hat{\mathbf{t}}$ (see Supplementary Fig.7), and $B$ is the magnitude of the magnetic field **B**. The magnetic field $\mathbf{B}(t)$ undergoes planar oscillations as given in (S1).

Formulation of the dynamic equations of the two-link microswimmer follows the same procedure as described above for the multi-link swimmer. Some differences are that there is no spherical link, and that the torque on the magnetic link is given by (S30) rather than (S3). Using this, one can obtain a system of differential equations of the form (S25), where the generalized coordinates are $\mathbf{q}=(x,y,\theta,\phi)$. An additional simplification is obtained if one replaces body velocity $\dot{x},\dot{y}$ by the components $v_t,v_n$ expressed in the body-fixed reference frame $\hat{\mathbf{t}},\hat{\mathbf{n}}$. The kinematic relation between these velocity components is given by:

$$\begin{pmatrix}\dot{x}\\ \dot{y}\end{pmatrix}=\begin{pmatrix}\cos\theta & -\sin\theta\\ \sin\theta & \cos\theta\end{pmatrix}\begin{pmatrix}v_t\\ v_n\end{pmatrix} \quad\text{(S31)}$$

The use of body-fixed velocity components $v_t,v_n$ eliminates the dependence of the left-hand side of (S25) on the absolute orientation angle $\theta$, resulting in considerable simplification, as:

$$\mathbf{A}(\phi)\begin{pmatrix}v_t\\ v_n\\ \dot{\theta}\\ \dot{\phi}\end{pmatrix}=\mathbf{b}(\theta,\phi,t) \quad\text{(S32)}$$

Where:

$$\mathbf{A}=c_t l\begin{pmatrix}\sin^2(\phi)+2 & \cos(\phi)\sin(\phi) & -\frac{1}{2}l(\cos(\phi)+2)\sin(\phi) & l\sin(\phi)\\ \cos(\phi)\sin(\phi) & \cos^2(\phi)+3 & -2l\cos^4\left(\frac{\phi}{2}\right) & l\cos(\phi)\\ -\frac{1}{2}l(\cos(\phi)+2)\sin(\phi) & -2l\cos^4\left(\frac{\phi}{2}\right) & \frac{1}{24}l^2(24\cos(\phi)+3\cos(2\phi)+29) & -\frac{1}{6}l^2(3\cos(\phi)+4)\\ l\sin(\phi) & l\cos(\phi) & -\frac{1}{6}l^2(3\cos(\phi)+4) & \frac{2l^2}{3}\end{pmatrix}$$

$$\mathbf{b}=\begin{pmatrix}0\\ 0\\ \Delta\chi v B_x^2(\cos(\theta)+\beta\sin(\theta)\sin(t\omega))(\beta\cos(\theta)\sin(t\omega)-\sin(\theta))\\ -k\phi\end{pmatrix}$$

After inversion of the matrix $\mathbf{A}$ in (S32), the dynamic equations of the two-link microswimmer are obtained as:

$$v_t = \frac{\sin(\phi)\cos(\phi)\left(\sin(2\theta)\left(\beta^2 \sin^2(2\pi ft)-1\right)+2\beta\cos(2\theta)\sin(2\pi ft)\right)}{2(\cos(2\phi)-17)}\frac{l}{t_m} - \frac{\phi\sin(\phi)(\cos(\phi)+3)}{2(\cos(2\phi)-17)}\frac{l}{t_k}$$

$$v_n = \frac{(5-\cos(2\phi))\left(\sin(2\theta)\left(-\beta^2+\beta^2\cos(4\pi ft)+2\right)-4\beta\cos(2\theta)\sin(2\pi ft)\right)}{16(\cos(2\phi)-17)}\frac{l}{t_m} + \frac{\phi\sin^2\left(\frac{\phi}{2}\right)(\cos(\phi)+3)}{2(\cos(2\phi)-17)}\frac{l}{t_k} \quad \text{(S33)}$$

$$\dot{\theta} = -\frac{(\cos(2\phi)+19)\left(\sin(2\theta)\left(\beta^2\sin^2(2\pi ft)-1\right)+2\beta\cos(2\theta)\sin(2\pi ft)\right)}{4(\cos(2\phi)-17)}\frac{1}{t_m} + \frac{\phi(\cos(\phi)+3)^2}{2(\cos(2\phi)-17)}\frac{1}{t_k}$$

$$\dot{\phi} = \frac{(\cos(\phi)+3)^2\left((\sin(2\theta)-\beta\sin(2\pi ft)(2\cos(2\theta)+\beta\sin(2\theta)\sin(2\pi ft)))\right)}{2(\cos(2\phi)-17)}\frac{1}{t_m} + \frac{\phi(\cos(\phi)+3)^2}{(\cos(2\phi)-17)}\frac{1}{t_k}$$

Where $t_m, t_k$ are characteristic time scales given by:

$$t_m = \frac{c_t l^2}{6\Delta\chi B_x^2 v} = \frac{viscous}{magnetic}, \quad t_k = \frac{c_t l^2}{12k} = \frac{viscous}{elastic} \quad \text{(S34)}$$

Note that (S33) is independent of the $x, y$ coordinates, since the fluid domain is unbounded. Therefore, the dynamic behavior of the swimmer is mostly encapsulated in the coupled nonlinear second-order system of the two angles $\theta, \phi$. We study it below using asymptotic analysis under three different physical limits.

**2.2.1 Asymptotic analysis under small-amplitude oscillations**

We now consider the dynamics in (S33) in the limit of small-amplitude oscillations of the magnetic field $\mathbf{B}(t)$ in (S1). By assuming $\theta \ll 1$, we can expand the solution into a power series in $\beta$ as:

$$\mathbf{q}(t) = \beta\mathbf{q}_1(t) + \beta^2\mathbf{q}_2(t) + \beta^3\mathbf{q}_3(t) + \ldots \quad \text{(S35)}$$

We expand the expressions for $\dot{\theta}, \dot{\phi}$ from (S33) into a Taylor series about $\theta=0, \phi=0$ and substitute (S35) into the expanded expressions. Next, we equate coefficients of same powers of $\beta$ in order to obtain a system of differential equations for each order of approximation. The first-order approximation yields the following system:

$$\begin{pmatrix}\dot{\theta}_1 \\ \dot{\phi}_1\end{pmatrix} = \begin{pmatrix} -\dfrac{5}{8t_m} & -\dfrac{1}{2t_k} \\ -\dfrac{1}{t_m} & -\dfrac{1}{t_k} \end{pmatrix}\begin{pmatrix}\theta_1 \\ \phi_1\end{pmatrix} + \begin{pmatrix}\dfrac{5}{8t_m} \\ \dfrac{1}{t_m}\end{pmatrix}\sin(\omega t) \quad \text{(S36)}$$

This is a linear system whose characteristic polynomial is:

$$\Delta(\lambda) = \lambda^2 + \left(\frac{5}{8t_m} + \frac{1}{t_k}\right)\lambda + \frac{1}{8t_m t_k}$$

Since $t_m, t_k$ are positive values, the system is stable. This implies that transient terms in the solution of (S36), which take the form $e^{-\lambda t}$, decay to zero in time. The steady-state solution of the system in (S36) contains harmonic terms which depend on the frequency $\omega$, and can be obtained as:

$$\theta_1(t) = \frac{t_k \omega^2 (25 t_k + 32 t_m) + 1}{\omega^2 \left(t_k^2 \left(64 t_m^2 \omega^2 + 25\right) + 64 t_k t_m + 64 t_m^2\right) + 1} \sin(\omega t) - \frac{8\omega \left(5 t_k^2 t_m \omega^2 + t_m\right)}{\omega^2 \left(t_k^2 \left(64 t_m^2 \omega^2 + 25\right) + 64 t_k t_m + 64 t_m^2\right) + 1} \cos(\omega t) \quad \text{(S37)}$$

$$\phi_1(t) = \frac{8 t_k \omega^2 (5 t_k + 8 t_m)}{\omega^2 \left(t_k^2 \left(64 t_m^2 \omega^2 + 25\right) + 64 t_k t_m + 64 t_m^2\right) + 1} \sin(\omega t) + \frac{8 t_k \omega \left(1 - 8 t_k t_m \omega^2\right)}{\omega^2 \left(t_k^2 \left(64 t_m^2 \omega^2 + 25\right) + 64 t_k t_m + 64 t_m^2\right) + 1} \cos(\omega t)$$

Next, it is possible to obtain the first-order solution for the body-fixed velocities $v_t, v_n$ in (S33) and expand the relation (S31) in $\theta$ in order to find the leading-order dynamics of forward motion $x(t)$, which is of order 2, and given by:

$$\dot{x}_2(t) = -\frac{\theta_1 \sin(\omega t)}{16 t_m} l - \frac{\phi_1 \sin(\omega t)}{16 t_m} l + \frac{\phi_1^2}{8 t_k} l + \frac{\theta_1^2}{16 t_m} l + \frac{\theta_1 \phi_1}{16 t_m} l \quad \text{(S38)}$$

Substituting the expressions for $\theta_1, \phi_1$ from (S37) into (S38), one obtains:

$$\dot{x}(t) = \left(V_x + C_x \sin(2\omega t) + D_x \cos(\omega t)\right) \beta^2 + O(\beta^4) \quad \text{(S39)}$$

Where

$$V_x = \frac{t_k \omega^2}{\omega^2 \left(t_k^2 \left(64 t_m^2 \omega^2 + 25\right) + 64 t_k t_m + 64 t_m^2\right) + 1} l$$

$$C_x = \frac{\omega \left(832 t_k^4 t_m^2 \omega^6 - t_k^2 \omega^4 \left(325 t_k^2 + 1408 t_k t_m + 1152 t_m^2\right) + 2\omega^2 \left(29 t_k^2 + 64 t_k t_m + 32 t_m^2\right) - 1\right)}{4 \left(\omega^2 \left(t_k^2 \left(64 t_m^2 \omega^2 + 25\right) + 64 t_k t_m + 64 t_m^2\right) + 1\right)^2} l$$

$$D_x = \frac{\omega^2 \left(4 t_k^3 t_m \omega^4 (65 t_k + 112 t_m) - t_k \omega^2 \left(45 t_k^2 + 120 t_k t_m + 64 t_m^2\right) + 3 t_k + 4 t_m\right)}{\left(\omega^2 \left(t_k^2 \left(64 t_m^2 \omega^2 + 25\right) + 64 t_k t_m + 64 t_m^2\right) + 1\right)^2} l$$

Note that the expression for the mean speed $V_x$ implies that it vanishes for extreme frequencies $\omega \to 0, \omega \to \infty$ and maximized for some intermediate optimal frequency $\omega_{opt}$ which can be obtained rather simply:

$$\omega_{opt} = \frac{l}{\sqrt{8t_m t_k}} \tag{S40}$$

Importantly, the small-$\beta$ approximation gives zero net motion in $y$ direction. Moreover, the leading-order system in (S36) is always stable, and expansion about the perpendicular orientation $\theta=\pi/2$, $\phi=0$, yields a linear system which is always *unstable*. Thus, the small-amplitude asymptotic approximation cannot capture the stability transitions and changes in the swimming direction from $x$ to $y$.

**2.2.2 Asymptotic analysis under fast oscillations**

We now analyze the two-link microswimmer's dynamics in the limit of fast oscillation frequency $\omega$ of the magnetic field, by using the method of multiple scales[7],[8]. We denote $\epsilon=(\omega t_m)^{-1}$, $\alpha\epsilon=(\omega t_k)^{-1}$ and assume that $\epsilon \ll 1$ and $\alpha=O(1)$. We define the nondimensional time $\tilde{t} = \omega t$, and introduce fast and slow time scales $T_0, T_1$ as:

$$T_0 = \tilde{t} = \omega t, \quad T_1 = \epsilon \tilde{t} = t/t_m$$

We expect to obtain solutions of fast oscillations superposed on slow evolution of "average" values, and so we assume solutions of the form: $\theta = \theta(T_0, T_1), \phi = \phi(T_0, T_1)$, and expand them into power series in $\epsilon$ as:

$$\theta = \theta_0(T_0, T_1) + \epsilon \theta_1(T_0, T_1) + \epsilon^2 \theta_2(T_0, T_1) + \ldots = \theta_0 + \Delta\theta$$
$$\phi = \phi_0(T_0, T_1) + \epsilon \phi_1(T_0, T_1) + \epsilon^2 \phi_2(T_0, T_1) + \ldots = \phi_0 + \Delta\phi \tag{S41}$$

From the chain rule, time-differentiation is expanded as:

$$\frac{d}{d\tilde{t}} = \frac{\partial}{\partial T_0} + \epsilon \frac{\partial}{\partial T_1}, \tag{S42}$$

and hence we define the time-derivative operators: $D_n = \frac{\partial}{\partial T_n}$ for $n=0,1$. Substituting (S41) and (S42) into the equations for $\theta, \phi$ in (S33), one obtains:

$$D_0\theta_0 + D_0\Delta\theta + \epsilon D_1\theta_0 + \epsilon D_1\Delta\theta = \frac{(\phi_0 + \Delta\phi)(\cos(\phi_0 + \Delta\phi) + 3)^2}{2(\cos(2(\phi_0 + \Delta\phi)) - 17)}\alpha\epsilon$$

$$-\frac{(\cos(2(\phi_0 + \Delta\phi)) + 19)\left(\sin(2(\theta_0 + \Delta\theta))(\beta^2\sin^2(T_0) - 1) + 2\beta\cos(2(\theta_0 + \Delta\theta))\sin(T_0)\right)}{4(\cos(2(\phi_0 + \Delta\phi)) - 17)}\epsilon$$

$$D_0\phi_0 + D_0\Delta\phi + \epsilon D_1\phi_0 + \epsilon D_1\Delta\phi = \frac{(\cos((\phi_0 + \Delta\phi)) + 3)^2(\phi_0 + \Delta\phi)}{(\cos(2(\phi_0 + \Delta\phi)) - 17)}\alpha\epsilon +$$

$$\frac{(\cos((\phi_0 + \Delta\phi)) + 3)^2\left((\sin(2(\theta_0 + \Delta\theta)) - \beta\sin(T_0)(2\cos(2(\theta_0 + \Delta\theta)) + \beta\sin(2(\theta_0 + \Delta\theta))\sin(T_0)))\right)}{2(\cos(2(\phi_0 + \Delta\phi)) - 17)}\epsilon$$

(S43)

We expand (S43) into a Taylor series in $\Delta\theta, \Delta\phi$ about 0 and equate coefficients of powers of $\epsilon$. Equating coefficients of order $\epsilon^0$ gives:

$$\begin{aligned} D_0\theta_0 &= 0 \\ D_0\phi_0 &= 0 \end{aligned} \quad \Rightarrow \quad \begin{aligned} \theta_0 &= \Theta_0(T_1) \\ \phi_0 &= \Phi_0(T_1) \end{aligned}$$

(S44)

That is, $\theta_0$ and $\phi_0$ are functions of the slow time $T_1$ only. Equating coefficients of order $\epsilon^1$ in the expansion of (S43) gives:

$$D_0\theta_1 + D_1\theta_0 = \frac{\left(4\alpha\phi_0(\cos(\phi_0) + 3)^2 + \sin(2\theta_0)(\cos(2\phi_0) + 19)(\beta^2\cos(2T_0) - \beta^2 + 2) - 4\beta\cos(2\theta_0)\sin(T_0)(\cos(2\phi_0) + 19)\right)}{8(\cos(2\phi_0) - 17)}$$

$$D_0\phi_1 + D_1\phi_0 = \frac{(\cos(\phi_0) + 3)^2\left(4\alpha\phi_0 + \sin(2\theta_0)(\beta^2\cos(2T_0) - \beta^2 + 2) - 4\beta\cos(2\theta_0)\sin(T_0)\right)}{4(\cos(2\phi_0) - 17)}$$

(S45)

Substituting the solutions for $\theta_0, \phi_0$ from (S44) into (S45) gives:

$$D_0\theta_1 + D_1\Theta_0 =$$
$$\frac{\left(4\alpha\Phi_0(\cos(\Phi_0) + 3)^2 + \sin(2\Theta_0)(\cos(2\Phi_0) + 19)(\beta^2\cos(2T_0) - \beta^2 + 2) - 4\beta\cos(2\Theta_0)\sin(T_0)(\cos(2\Phi_0) + 19)\right)}{8(\cos(2\Phi_0) - 17)}$$

$$D_0\phi_1 + D_1\Phi_0 = \frac{(\cos(\Phi_0) + 3)^2\left(4\alpha\Phi_0 + \sin(2\Theta_0)(\beta^2\cos(2T_0) - \beta^2 + 2) - 4\beta\cos(2\Theta_0)\sin(T_0)\right)}{4(\cos(2\Phi_0) - 17)}$$

(S46)

Integrating (S46) with respect to the fast time $T_0$ gives:

$$\theta_1(T_0, T_1) = \frac{\beta^2\sin(2\Theta_0)(\cos(2\Phi_0) + 19)}{16(\cos(2\Phi_0) - 17)}\sin(2T_0) + \frac{\beta\cos(2\Theta_0)(\cos(2\Phi_0) + 19)}{2(\cos(2\Phi_0) - 17)}\cos(T_0)$$

$$+\frac{\left(-2(\beta^2 - 2)\sin(2\Theta_0)(\cos(2\Phi_0) + 19) - 8(\cos(\Phi_0) + 3)(\cos(\Phi_0)(4D_1\Theta_0 - \alpha\Phi_0) - 3(4D_1\Theta_0 + \alpha\Phi_0))\right)}{16(\cos(2\Phi_0) - 17)}T_0 + \Theta_1(T_1)$$

$$\phi_1\theta_1(T_0, T_1) = \frac{\beta^2\sin(\Theta_0)\cos(\Theta_0)(\cos(\Phi_0) + 3)}{8(\cos(\Phi_0) - 3)}\sin(2T_0) + \frac{\beta\cos(2\Theta_0)(\cos(\Phi_0) + 3)}{2(\cos(\Phi_0) - 3)}\cos(T_0) +$$

$$\frac{\left(4(3\alpha\Phi_0 + \cos(\Phi_0)(\alpha\Phi_0 - 2D_1\Phi_0) + 6D_1\Phi_0) - 2(\beta^2 - 2)\sin(\Theta_0)\cos(\Theta_0)(\cos(\Phi_0) + 3)\right)}{8(\cos(\Phi_0) - 3)}T_0 + \Phi_1(T_1)$$

(S47)

Since we assume an oscillating solution in the fast time $T_0$, we require that the diverging secular terms (multiplied by $T_0$) in (S47) are eliminated. We obtain the following system for the zero-order approximation of the slow dynamics of $\theta, \phi$:

$$\frac{d\Theta_0}{dT_1} = \frac{4\alpha\Phi_0(\cos(\Phi_0)+3)^2 - (\beta^2-2)\sin(2\Theta_0)(\cos(2\Phi_0)+19)}{8(\cos(2\Phi_0)-17)}$$
$$\frac{d\Phi_0}{dT_1} = \frac{(\cos(\Phi_0)+3)\left(4\alpha\Phi_0 - (\beta^2-2)\sin(2\Theta_0)\right)}{8(\cos(\Phi_0)-3)} \tag{S48}$$

This is a second-order coupled nonlinear system of time-independent differential equations in the slow time $T_1$. This system has two equilibrium points: $\{\theta_e = 0, \phi_e = 0\}$ and $\{\theta_e = \frac{\pi}{2}, \phi_e = 0\}$ (more precisely, $\theta_e = \pm k\frac{\pi}{2}$ for $k = 0,1,2\ldots$ but the all other value are ignored due to symmetry). Linearization about $\Theta_0 = \theta_e$ yields:

$$\begin{pmatrix} D_1\Theta_0 \\ D_1\Phi_0 \end{pmatrix} = \begin{pmatrix} \frac{5}{16}(\beta^2-2)\cos(2\theta_e) & -\frac{1}{2}\alpha \\ \frac{1}{2}(\beta^2-2)\cos(2\theta_e) & -\alpha \end{pmatrix} \begin{pmatrix} \Theta_0 \\ \Phi_0 \end{pmatrix} \tag{S49}$$

The characteristic polynomial for this system is:

$$\lambda^2 + \left(\alpha - \frac{5}{16}(\beta^2-2)\cos(2\theta_e)\right)\lambda - \frac{1}{16}\alpha(\beta^2-2)\cos(2\theta_e)$$

In order for a 2$^{\text{nd}}$ order system to be locally asymptotically stable all the coefficients of its characteristic polynomial are required to be positive, hence for $\theta_e = 0$ the system is stable when $\beta < \sqrt{2}$, while for $\theta_e = \frac{\pi}{2}$ the system is stable when $\beta > \sqrt{2}$.

Therefore, steady-state solutions for $\Theta_0, \Phi_0$ depend on the value of $\beta$ as:

$$\Phi_0 = 0, \quad \Theta_0 = \begin{cases} 0 & \text{for } \beta < \sqrt{2} \\ \frac{\pi}{2} & \text{for } \beta < \sqrt{2} \end{cases} \tag{S50}$$

Next, we look for approximate solutions of higher orders for $\theta, \phi$ in steady state for oscillations about either $\theta_e = 0$ or $\theta_e = \dfrac{\pi}{2}$. Substituting the steady state value of $\Theta_0, \Phi_0$ (S50) into (S47) yields:

$$\theta_1 = \Theta_1 - \frac{5}{8}\beta\cos(2\theta_e)\cos(\tau)$$
$$\phi_1 = \Phi_1 - \beta\cos(2\theta_e)\cos(\tau)$$
(S51)

From (S33), the next-order approximation is:

$$D_0\theta_2 = \frac{5}{128}\cos(2\theta_e)\left(\beta^2\cos(2T_0)-\beta^2+2\right)(5\beta\cos(2\theta_e)\cos(T_0)-8\Theta_1) - \frac{1}{2}\alpha(\Phi_1 - \beta\cos(2\theta_e)\cos(T_0))$$

$$D_0\phi_2 = \alpha\beta\cos(2\theta_e)\cos(T_0) - \alpha\Phi_1 + \frac{1}{16}\cos(2\theta_e)\left(\beta^2\cos(2\tau)-\beta^2+2\right)(5\beta\cos(2\theta_e)\cos(T_0)-8\Theta_1)$$

Integrating in $T_0$ gives

$$\theta_2 = \sin(T_0)\left(\frac{1}{2}\alpha\beta\cos(2\theta_e) - \frac{25}{256}\beta\left(\beta^2-4\right)\cos^2(2\theta_e)\right) + \frac{25}{768}\beta^3\cos^2(2\theta_e)\sin(3T_0)$$
$$-\frac{5}{32}\beta^2\Theta_1\cos(2\theta_e)\sin(2T_0) + \left(-\frac{\alpha\Phi_1}{2} + \frac{5}{16}(\beta^2-2)\Theta_1\cos(2\theta_e) - D_1\Theta_1\right)T_0 + \Theta_2(T_1)$$

$$\phi_2 = \sin(T_0)\left(\alpha\beta\cos(2\theta_e) - \frac{5}{32}\beta\left(\beta^2-4\right)\cos^2(2\theta_e)\right) + \frac{5}{96}\beta^3\cos^2(2\theta_e)\sin(3T_0)$$
$$-\frac{1}{4}\beta^2\Theta_1\cos(2\theta_e)\sin(2T_0) + \left(-\alpha\Phi_1 + \frac{1}{2}(\beta^2-2)\Theta_1\cos(2\theta_e) - D_1\Phi_1\right)T_0 + \Phi_2(T_1)$$
(S52)

Requiring elimination of (diverging) secular terms yields:

$$\begin{pmatrix} D_1\Theta_1 \\ D_1\Phi_1 \end{pmatrix} = \begin{pmatrix} \dfrac{5}{16}(\beta^2-2)\cos(2\theta_e) & -\dfrac{1}{2}\alpha \\ \dfrac{1}{2}(\beta^2-2)\cos(2\theta_e) & -\alpha \end{pmatrix}\begin{pmatrix} \Theta_1 \\ \Phi_1 \end{pmatrix}$$
(S53)

This system (S53) is similar to the system in (S49) and has the same stability conditions. The steady-state solution of the system (S49) for a stable point $\theta_e$ converges to zero. Substituting the steady state solution $\Theta_1 = \Phi_1 = 0$ into (S52) yields:

$$\theta_2 = \sin(\tau)\left(\frac{1}{2}\alpha\beta\cos(2\theta_e) - \frac{25}{256}\beta\left(\beta^2-4\right)\cos^2(2\theta_e)\right) + \frac{25}{768}\beta^3\cos^2(2\theta_e)\sin(3\tau) + \Theta_2(T_1)$$
$$\phi_2 = \sin(\tau)\left(\alpha\beta\cos(2\theta_e) - \frac{5}{32}\beta\left(\beta^2-4\right)\cos^2(2\theta_e)\right) + \frac{5}{96}\beta^3\cos^2(2\theta_e)\sin(3\tau) + \Phi_2(T_1)$$
(S54)

Further expansions of the solutions to higher orders is done iteratively in a similar manner, up to any desired order in $\epsilon$. The approximate steady-state solution is then obtained by substituting the steady-state solution of each order into (S41):

$$\theta \approx \theta_e + \epsilon\theta_1(T_0,\theta_e) + \epsilon^2\theta_2(T_0,\theta_e) + \ldots + O(\epsilon^n)$$
$$\phi \approx \epsilon\phi_1(T_0,\theta_e) + \epsilon^2\phi_2(T_0,\theta_e) + \ldots + O(\epsilon^n) \tag{S55}$$

Next, the expressions in (S55) are substituted into (S33) in order to obtain leading-order expressions for the body velocity components $v_t, v_n$. These body-frame velocities are transformed into world-frame velocities $\dot{x}, \dot{y}$ by using the relation (S31), expanding it as a power series in $(\theta, \phi)$ about $(\theta_e, 0)$, and substituting (S55). The leading-order expressions for $\dot{x}, \dot{y}$ contain oscillating terms and constant terms $V_x, V_y$, which are precisely the mean speed components, given by:

$$V_x = (\cos(\theta_e) + \cos(3\theta_e))\left(\frac{1}{128}\alpha\beta^2\epsilon^3 + \frac{\alpha\beta^2\left(-9(77-384\alpha)\beta^2 - 72(64\alpha(\alpha+1)+25) - 625\beta^4\right)}{589824}\epsilon^5\right)$$

$$V_y = (\sin(\theta_e) - \sin(3\theta_e))\left(\frac{1}{128}\alpha\beta^2\epsilon^3 + \frac{\alpha\beta^2\left(-9(384\alpha+77)\beta^2 - 72(64(\alpha-1)\alpha+25) - 625\beta^4\right)}{589824}\epsilon^5\right) \tag{S56}$$

It can be seen from (S56) that in the case where oscillations about $\theta_e = 0$ are stable $\left(\beta < \sqrt{2}\right)$ the swimmer moves in $\hat{\mathbf{x}}$ direction, $V_x \neq 0$, $V_y = 0$. On the other hand, in the case where oscillations about $\theta_e = \frac{\pi}{2}$ are stable $\left(\beta > \sqrt{2}\right)$ the swimmer moves in $\hat{\mathbf{y}}$ direction, $V_x = 0$, $V_y \neq 0$. Finally, the expressions in (S56) imply the existence of an optimal value of $\beta$ that maximizes the speed $V_y$ in the regime of $\beta > \sqrt{2}$.

### 2.2.3 Asymptotic analysis under a stiff spring

We now consider the limit where the torsion spring is very stiff and oscillations frequency are very fast. This implies that $\epsilon=(\omega t_m)^{-1}$, where $\epsilon \ll 1$, while $(\omega t_k)^{-1}=O(1)$. We use the same nondimensional time $\tilde{t} = \omega t$, and expand the joint angle $\phi$ into a power series in $\epsilon$ as:

$$\phi(t) = \phi_0(t) + \epsilon \phi_1(t) + \epsilon^2 \phi_2(t) + \ldots = \phi_0(t) + \Delta\phi(t) \tag{S57}$$

Substituting (S57) into the system (S33), the equation for $\dot{\theta}$ is obtained as:

$$\dot{\theta} = \frac{\phi_0(\cos(\phi_0)+3)^2}{2(\cos(2\phi_0)-17)}\frac{1}{t_k\omega} + \epsilon\left(\frac{\phi_1(\cos(\phi_0)+3)^2(12\phi_0\sin(\phi_0)+\cos(2\phi_0)-17)}{2(\cos(2\phi_0)-17)^2}\alpha_k + \frac{(\cos(2\phi_0)+19)\left(\sin(2\theta)\left(\beta^2\cos(2\tau)-\beta^2+2\right)-4\beta\cos(2\theta)\sin(\tau)\right)}{8(\cos(2\phi_0)-17)}\right) + O(\epsilon^2) \tag{S58}$$

For the equation in $\phi$ from (S33), we equate expressions in different of powers of $\epsilon$:

$$\epsilon^0 : \dot{\phi}_0 = -\frac{\phi_0(\cos(\phi_0)+3)^2}{17-\cos(2\phi_0)}\frac{1}{t_k\omega} = -f_0(\phi_0)\phi_0$$

$$\epsilon^1 : \dot{\phi}_1 = \frac{\phi_1(12\phi_0\sin(\phi_0)+\cos(2\phi_0)-17)}{4(\cos(\phi_0)-3)^2}\frac{1}{t_k\omega} - \frac{(\cos(\phi_0)+3)^2\left(\sin(2\theta)\left(\beta^2\cos(2\tau)-\beta^2+2\right)-4\beta\cos(2\theta)\sin(\tau)\right)}{4(17-\cos(2\phi_0))} \tag{S59}$$

It can easily be seen that the solution for $\phi_0(t)$ decays asymptotically to zero, since the function $f_0(\phi_0)$ in (S59) is bounded between two positive constants. Thus, in steady-state we can use the first-order approximation $\phi = \epsilon\phi_1$. Substituting into (S33) and re-normalizing time and frequency by $t_m$, one obtains the system:

$$\dot{\theta} = -\frac{1}{2}\alpha\phi - \frac{5}{32}\sin(2\theta)\left(\beta^2\cos(2t\omega) - \beta^2 + 2\right) + \frac{5}{8}\beta\cos(2\theta)\sin(t\omega)$$

$$\dot{\phi} = -\alpha\phi - \frac{1}{4}\sin(2\theta)\left(\beta^2\cos(2t\omega) - \beta^2 + 2\right) + \beta\cos(2\theta)\sin(t\omega) \tag{S60}$$

The system (S60) is nonlinear in $\theta(t)$ while $\phi(t)$ appears linearly. Using the time-differentiation operator $D$, we substitute $\dot{\phi} = D\phi$ into (S60) and eliminate $\phi$ in order to obtain a single equation for $\theta$ as:

$$(D+\alpha)\dot{\theta} = -\frac{1}{2}\alpha\left(-\frac{1}{4}\sin(2\theta)\left(\beta^2\cos(2t\omega)-\beta^2+2\right)+\beta\cos(2\theta)\sin(t\omega)\right)$$

$$-(D+\alpha)\frac{5}{32}\sin(2\theta)\left(\beta^2\cos(2t\omega)-\beta^2+2\right)+(D+\alpha)\frac{5}{8}\beta\cos(2\theta)\sin(t\omega)$$

This equation can be rearranged as a second-order Kapitza-like differential equation:

$$32\ddot{\theta}+\left(32\alpha+10\cos(2\theta)\left(\beta^2\cos(2\omega t)-\beta^2+2\right)+40\beta\sin(2\theta)\sin(\omega t)\right)\dot{\theta}$$

$$+\left(\alpha(2-\beta^2)+\alpha\beta^2\cos(2\omega t)-10\omega\beta^2\sin(2\omega t)\right)\sin(2\theta)$$

$$=4(5\omega\beta\cos(\omega t)+\alpha\beta\sin(\omega t))\cos(2\theta) \quad \text{(S61)}$$

**2.2.4 Variational equation and stability of periodic solutions**

The nonlinear system in S(61) typically has a periodic solution of nonlinear oscillations $\theta_p(t) = \theta_p(t+T)$, where $T = 2\pi/\omega.$, We now study the stability of this periodic solution by considering a small perturbation about it:

$$\theta(t) = \theta_p(t) + \delta(t) \quad \text{(S62)}$$

Substituting (S62) into S(61), expanding trigonometric expressions and rearranging gives:

$$\ddot{\delta}+\frac{1}{8}\left(8\alpha+5\cos(2\theta_p)\left(1-\beta^2\sin(\omega t)^2\right)+10\beta\sin(2\theta_p)\sin(t\omega)\right)\dot{\delta}$$

$$+\frac{1}{8}\begin{pmatrix}10\left(\sin(2\theta_p)\left(\beta^2\sin^2(t\omega)-1\right)+2\beta\cos(2\theta_p)\sin(t\omega)\right)\dot{\theta}_p\\+\cos(2\theta_p)\left(\alpha\left(1-\beta^2\sin(\omega t)^2\right)-5\beta^2\omega\sin(2t\omega)\right)\\+2\beta\sin(2\theta_p)(\alpha\sin(t\omega)+5\omega\cos(t\omega))\end{pmatrix}\delta$$

$$+\begin{bmatrix}\ddot{\theta}_p+\left(\alpha+\frac{5}{16}\cos(2\theta_p)\left(\beta^2\cos(2t\omega)-\beta^2+2\right)+\frac{5}{4}\beta\sin(2\theta_p)\sin(t\omega)\right)\dot{\theta}_p\\+\left(\frac{1}{32}\alpha\left(\beta^2\cos(2t\omega)-\beta^2+2\right)-\frac{5}{16}\omega\beta^2\sin(2t\omega)\right)\sin(2\theta_p)\\+\left(-\frac{1}{8}\alpha\beta\sin(t\omega)-\frac{5}{8}\beta\omega\cos(t\omega)\right)\cos(2\theta_p)\end{bmatrix} \quad \text{(S63)}$$

$$+\left\{-\frac{5}{4}\left(\sin(2\theta_p)\left(\beta^2\sin^2(t\omega)-1\right)+2\beta\cos(2\theta_p)\sin(t\omega)\right)\delta\dot{\delta}\right\}=0$$

The terms in the square brackets in (S63) vanish since $\theta_p(t)$ is a solution of S(61). Assuming small perturbation $\delta(t)$, the higher-order terms appearing in braces in (S63) also vanish, and we are left with the variational equation:

$$\ddot{\delta} + \frac{1}{8}\left(8\alpha + 5\cos(2\theta_p)\left(1 - \beta^2\sin(\omega t)^2\right) + 10\beta\sin(2\theta_p)\sin(t\omega)\right)\dot{\delta}$$

$$+ \frac{1}{8}\begin{pmatrix} 10\left(\sin(2\theta_p)\left(\beta^2\sin^2(t\omega)-1\right) + 2\beta\cos(2\theta_p)\sin(t\omega)\right)\dot{\theta}_p \\ +\cos(2\theta_p)\left(\alpha\left(1-\beta^2\sin(\omega t)^2\right) - 5\beta^2\omega\sin(2t\omega)\right) \\ +2\beta\sin(2\theta_p)(\alpha\sin(t\omega) + 5\omega\cos(t\omega)) \end{pmatrix}\delta = 0 \quad \text{(S64)}$$

This is a Hill-type equation[8], a second-order linear homogenous differential equation with time-periodic coefficients[8]. According to Floquet theory[8], this equation has two fundamental solutions $\delta_1(t), \delta_2(t)$, which satisfy $\delta_i(t+T) = \lambda_i \delta_i(t)$ for $i = 1, 2$, where $\lambda_i$ are called "Floquet multipliers". The conditions for local orbital stability of the periodic solution $\theta_p(t)$ is thus reduced to the requirement that $|\lambda_i| < 1$ for $i = 1, 2$, which implies that any small perturbation $\delta(t)$ decays asymptotically to zero. The Floquet multipliers can be found numerically as follows. The linearity of (S64) implies that its solution satisfies

$$\mathbf{z}(t+T) = \mathbf{P}\mathbf{z}(t), \text{ where } \mathbf{z} = \begin{pmatrix} \delta \\ \dot{\delta} \end{pmatrix}. \quad \text{(S65)}$$

Columns of the matrix $\mathbf{P}$ can be obtained from solutions of $\mathbf{z}(T)$ under initial conditions of $\mathbf{z}(0) = (1 \ 0)^T$ and $\mathbf{z}(0) = (0 \ 1)^T$. The Floquet multipliers are simply the eigenvalues of $\mathbf{P}$. Thus, the stability of a periodic solution can be obtained by numerical integration of (S64) over a single time period $T$, once the periodic solution $\theta_p(t)$ of S(61) is found numerically.

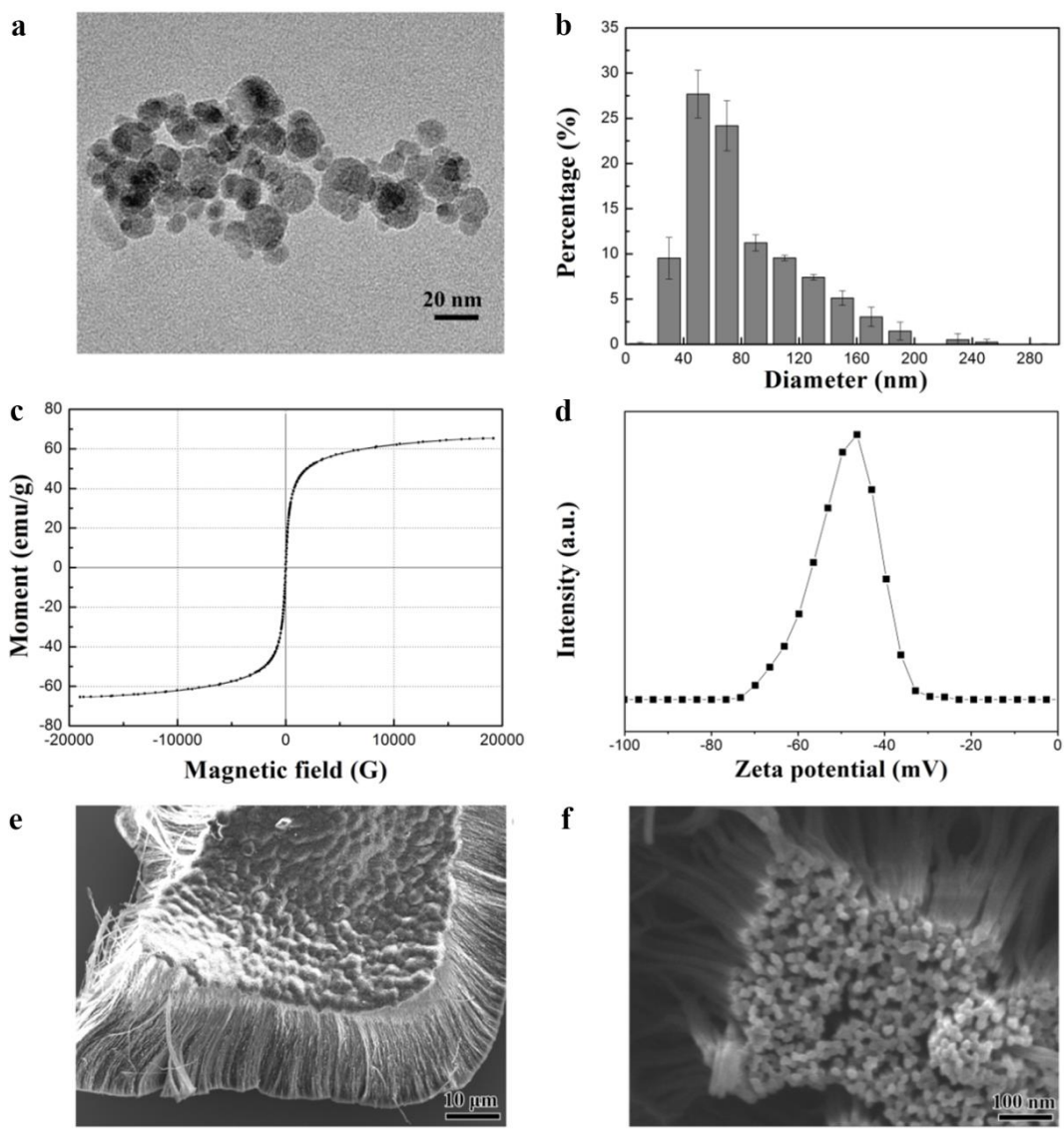

**Supplementary Figure 1|Characterizations of $Fe_3O_4$ nanoparticles and microswimmers.**
**a**, TEM micrograph of $Fe_3O_4$ nanoparticles. **b**, Size distribution of $Fe_3O_4$ nanoparticles. The standard deviation values were obtained by repeating the measurement for three times. **c**, Magnetic hysteresis loop of $Fe_3O_4$ nanoparticles. **d**, Surface charge distribution of $Fe_3O_4$ nanoparticles. **e-f**, SEM images of the soft microswimmers.

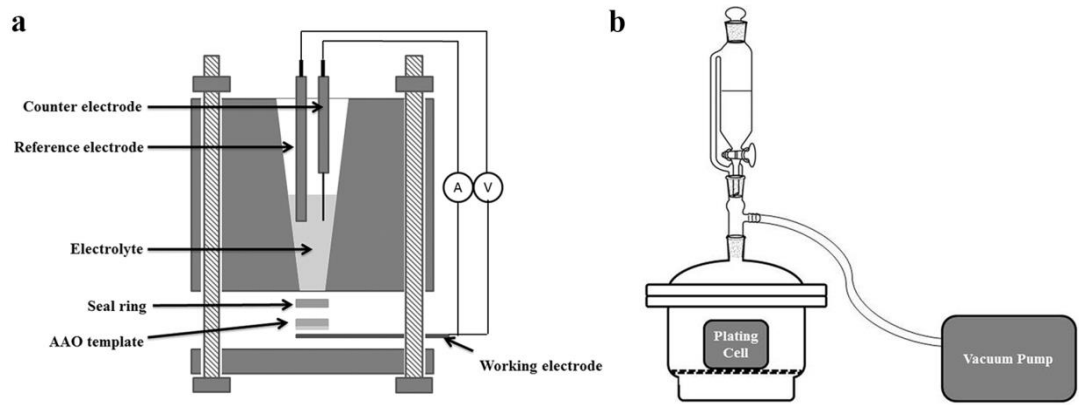

**Supplementary Figure 2|Illustration of the homemade teflon plating cell (a) and modified vacuum drier (b).**

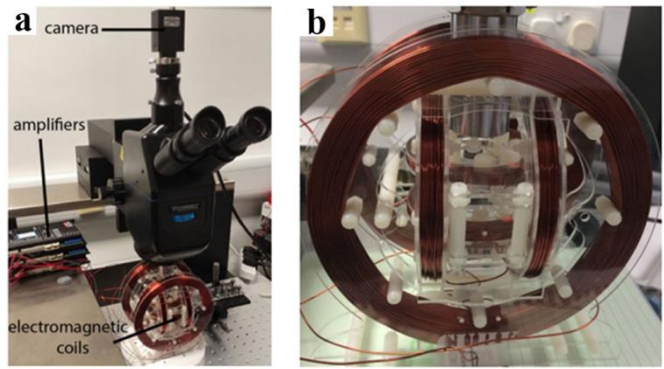

**Supplementary Figure 3|Setup of the dynamic magnetic field generation and actuation system, which consists of three pairs of Helmholtz electromagnetic coils assembled in an orthogonal manner**

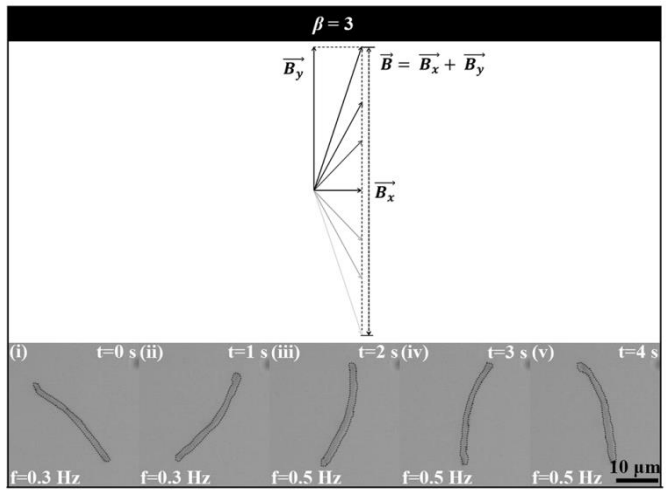

**Supplementary Figure 4|The 90 °-transition in oscillating direction of microswimmer when *β* was 3 and *f* was increased from 0.3 to 0.5 Hz.** Movie clips are available in the Supplementary Movie 4.

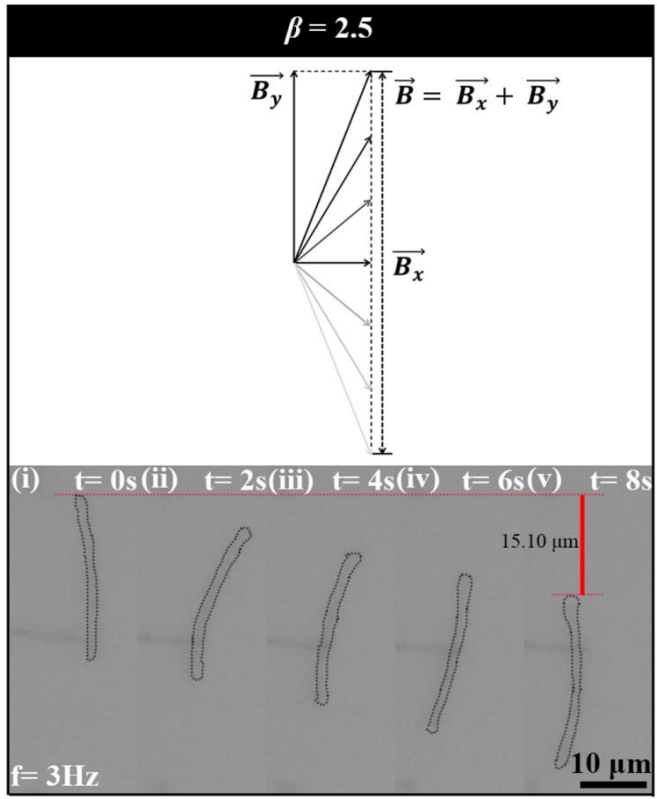

**Supplementary Figure 5|The dynamic motion of microswimmer in y direction when *β* was 2.5 and *f* was 3 Hz.** Full movies were enclosed in Supplementary Movie 5.

**Supplementary Figure 6|Schematic model of the multi-link swimmer.**

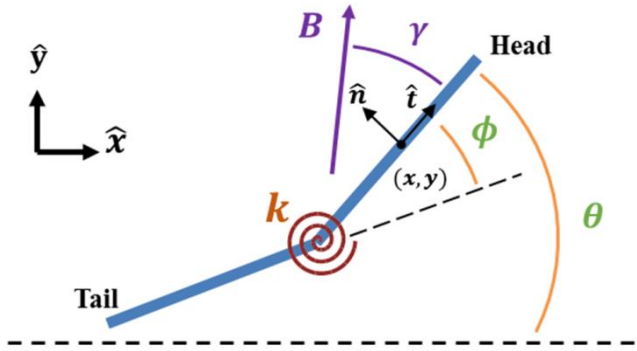

**Supplementary Figure 7|Model of the two-link microswimmer.**

**Supplementary Table 4 | List of parameters used for numerical analysis, taken from Dreyfus *et al*[1]**

| PARAMETER | MEANING | GREEN SQUARES ■ | BLUE DIAMONDS ♦ | RED CIRCLES ● |
|---|---|---|---|---|
| $\kappa$ | Flexural rigidity | $3.3 \cdot 10^{-22} \left[\frac{J}{m}\right]$ | $3.3 \cdot 10^{-22} \left[\frac{J}{m}\right]$ | $3.3 \cdot 10^{-22} \left[\frac{J}{m}\right]$ |
| $L$ | Filament length | $13 [\mu m]$ | $21 [\mu m]$ | $24 [\mu m]$ |
| $B_x$ | Constant field component | $8.7 [mT]$ | $8.7 [mT]$ | $8.9 [mT]$ |
| $B_y$ | Oscillating field component amplitude | $9.3 [mT]$ | $9.3 [mT]$ | $10.3 [mT]$ |
| $a_s$ | Blood cell radius | $3.2 [\mu m]$ | $2.7 [\mu m]$ | $3.1 [\mu m]$ |
| $\chi_=$ | Susceptibility in the preferred direction | 1 | 1 | 1 |
| $\chi_\perp$ | Susceptibility in the perpendicular direction | 1 | 1 | 1 |
| $c_n$ | Normal drag coefficient | $4\pi\mu$ | $4\pi\mu$ | $4\pi\mu$ |
| $a$ | Bead radius | $0.5 [\mu m]$ | $0.5 [\mu m]$ | $0.5 [\mu m]$ |